\documentclass[draftcls, onecolumn]{IEEEtran}
\ifCLASSINFOpdf
\else
\fi
\usepackage{graphicx}
\usepackage{amsmath}
\usepackage{amssymb}
\usepackage{graphics}
\usepackage{latexsym}
\usepackage[dvips]{epsfig}
\usepackage[nospace]{cite}
\usepackage{subfigure}
\usepackage{setspace}
\usepackage{tabularx}

\newtheorem{Definition}{Definition}
\newtheorem{Lemma}{Lemma}

\newtheorem{Proposition}[Lemma]{Proposition}
\newtheorem{Theorem}{\bf \emph{Theorem}}

\newtheorem{Example}{Example}
\newtheorem{Remark}{Remark}

\allowdisplaybreaks[1]
\usepackage[ colorlinks = true,
             linkcolor = blue,
             urlcolor  = blue,
             citecolor = red,
             anchorcolor = green,
]{hyperref}

\def\Pr{{\mathrm{Pr}}}
\def\E{{\mathrm E}}
\def\Var{{\mathrm {Var}}}

\begin{document}
%
\title{\huge Cut-Set Bounds for Networks with Zero-Delay Nodes}
%
%
%

\author{Silas~L.~Fong$^\dagger$ and Raymond~W.~Yeung$^*$
  \thanks{$^\dagger$ S.~L.~Fong is with the Department of Electrical and Computer Engineering, National University of Singapore, Singapore (e-mail: \texttt{silas\_fong@nus.edu.sg})}.
     \thanks{$^*$ R.~W.~Yeung is with the Institute of Network Coding and the Department
of Information Engineering, The Chinese University of Hong Kong, Hong Kong (e-mail: \texttt{whyeung@ie.cuhk.edu.hk}). He is also with the Shenzhen Key Laboratory of Network
    Coding Key Technology and Application, and Shenzhen Research
    Institute, The Chinese University of Hong Kong, Shenzhen, China. }
\thanks{This paper was presented in part at \textit{IEEE ISIT'12}, Jul. 2012.}}
\maketitle

\begin{abstract}
In a network, a node is said to \textit{incur a delay} if its encoding of each transmitted symbol involves only its received symbols obtained before the time slot in which the transmitted symbol is sent (hence the transmitted symbol sent in a time slot cannot depend on the received symbol obtained in the same time slot). A node is said to \textit{incur no delay} if its received symbol obtained in a time slot is available for encoding its transmitted symbol sent in the same time slot. Under the classical model, every node in a discrete memoryless network (DMN) incurs a unit delay, and the capacity region of the DMN satisfies the well-known cut-set outer bound. In this paper, we propose a generalized model for the DMN where some nodes may incur no delay. Under our generalized model, we obtain a new cut-set outer bound, which is proved to be tight for some two-node DMN and is shown to subsume an existing cut-set bound for the causal relay network \cite{causalRelayNetwork}. In addition, we establish under the generalized model another cut-set outer bound on the \textit{positive-delay region} -- the set of achievable rate tuples under the constraint that every node incurs a delay. We use the cut-set bound on the positive-delay region to show that for some two-node DMN under the generalized model, the positive-delay region is strictly smaller than the capacity region.
\end{abstract}


%
\IEEEpeerreviewmaketitle

\section{Introduction}
%
%
%
%
\IEEEPARstart{T}{his} paper considers a general network in which each node may send information to the other nodes. A node is said to \textit{incur a delay} if its encoding of each transmitted symbol involves only its received symbols obtained before the time slot in which the transmitted symbol is sent. A node is said to \textit{incur no delay} if its received symbol obtained in a time slot is available for encoding its transmitted symbol sent in the same time slot. In the classical model of the discrete memoryless network (DMN) \cite[Chapter 15]{CoverBook}, every node incurs a unit delay. We call the DMN under the classical model the \textit{classical DMN}. A well-known result for the classical DMN is the cut-set outer bound \cite{AbbasCutset,CoverBook}. However, the delay assumption makes the classical model not applicable to some simple networks including the \textit{relay-without-delay channel} studied by El~Gamal \textit{et al}.\ \cite[Section IV]{AbbasRelayNetwork} and the \textit{causal relay network} by Baik and Chung \cite{causalRelayNetwork}, where the causal relay network is a generalization of the relay-without-delay channel. Therefore, we are motivated to generalize the model of the DMN in such a way that some nodes may incur no delay. In our generalized model, some nodes may incur no delay, and we call the DMN under the generalized model the \textit{generalized DMN}. We are not only interested in the capacity region of the generalized DMN, but also the set of achievable rate tuples under the constraint that every node incurs a delay, and we call the constrained achievable rate region the \textit{positive-delay region}.
\subsection{Main Contribution}
In this paper, we prove a new cut-set bound on the capacity region and another cut-set bound on the positive-delay region under the generalized model. Then, we characterize the capacity region of some two-node generalized DMN by demonstrating an optimal transmission scheme that achieves our cut-set bound on the capacity region. In addition, we use the cut-set bound on the positive-delay region to show that for the two-node generalized DMN, the positive-delay region is strictly smaller than the capacity region, and hence some rate tuples in the capacity region cannot be achieved by imposing the classical constraint that every node must incur a delay.

On the other hand, we apply our cut-set bounds in the well-studied causal relay network in \cite{causalRelayNetwork}. We use our cut-set bound on the capacity region to recover an existing cut-set bound for the causal relay network. In addition, we use our cut-set bound on the positive-delay region to show that for some Gaussian causal relay network, the positive-delay region is strictly smaller than the capacity region.

One important consequence of our work is the following statement: Under the generalized model, the set of achievable rate tuples for a given DMN may depend on the amount of non-negative delay incurred by each node. The above statement complements the work by Effros \cite{EffrosIndependentDelay}, which shows that under the classical model, the set of achievable rate tuples for any DMN does not depend on the amount of positive delay incurred by each node.


\subsection{Paper Outline}
This paper is organized as follows. Section~\ref{notation} presents the notation of this paper. Section~\ref{motivatingExamples} presents a two-node network in which the classical cut-set bound cannot be applied. Section~\ref{sectionDefinition} presents the formulation of the generalized DMN. Section~\ref{sectionCapacityRegions} defines the capacity region and the positive-delay region of the generalized DMN and states our main results. Section~\ref{sectionOuterBound} proves the cut-set bound on the capacity region of the generalized DMN. Section~\ref{sectionGeneralized2way} investigates a two-node generalized DMN where one node incurs no delay. Our cut-set bound is shown to be tight for the two-node generalized DMN and hence the capacity region is determined.
Section~\ref{sectionOuterBound*} proves the cut-set bound on the positive-delay region. Section~\ref{illustratingExample} applies our cut-set bound on the positive-delay region to a two-node generalized DMN and shows that the positive-delay region is strictly smaller than the capacity region. Section~\ref{sectionRelayWithoutDelay} demonstrates that our cut-set bound on the capacity region subsumes an existing cut-set bound for the causal relay network. In addition, we use our cut-set bound on the positive-delay region to show that for some Gaussian causal relay network, the positive-delay region is strictly smaller than the capacity region.
Section~\ref{conclusion} concludes this paper.

\section{Notation}\label{notation}
We use $\Pr\{\mathcal{E}\}$ to represent the probability of an
event~$\mathcal{E}$. We use a capital letter~$X$ to denote a random variable with alphabet $\mathcal{X}$, and use the small letter $x$ to denote the realization of~$X$.
We use $X^n$ to denote a random tuple $(X_1,  X_2,  \ldots,  X_n)$, where the components $X_k$ have the same alphabet~$\mathcal{X}$.
 We let $p_X$ and $p_{Y|X}$ denote the probability mass distribution of $X$ and the conditional probability mass distribution of $Y$ given $X$ respectively for any discrete random variables~$X$ and~$Y$. We let $p_X(x)\triangleq Pr\{X=x\}$ and $p_{Y|X}(y|x)\triangleq Pr\{Y=y|X=x\}$ be the evaluations of $p_X$ and $p_{Y|X}$ respectively at $X=x$ and $Y=y$. We let $p_Xp_{Y|X}$ denote the joint distribution of $(X,Y)$, i.e., $p_Xp_{Y|X}(x,y)=p_X(x)p_{Y|X}(y|x)$ for all $x$ and $y$. If $X$ and $Y$ are independent, their joint distribution is simply $p_X p_Y$. We will take all logarithms to the base 2.
For any discrete random variables $(X,Y,Z)$ distributed according to $p_{X,Y,Z}$, we let $H_{p_{X,Z}}(X|Z)$ and $I_{p_{X,Y,Z}}(X;Y|Z)$ be the entropy of $X$ given $Z$ and mutual information between $X$ and $Y$ given $Z$  respectively. For simplicity, we drop the subscript in a notation if there is no ambiguity.
  If $X$, $Y$ and $Z$ are distributed according to $p_{X,Y,Z}$ and they form a Markov chain, we write
$(X\rightarrow Y\rightarrow Z)_{p_{X,Y,Z}}$ or more simply, $(X\rightarrow Y\rightarrow Z)_p$. The sets of natural and real numbers are denoted by $\mathbb{N}$ and $\mathbb{R}$ respectively.
For any $N^2$-dimensional random tuple
\[
(W_{1,1}, W_{1,2}, \ldots, W_{N,N}) \in \mathcal{W}_{1,1}\times \mathcal{W}_{1,2} \times \ldots \times \mathcal{W}_{N,N}
\]
 and any
set $V\subseteq \{1, 2, \ldots, N\}^2$, we let
\[
W_V=(W_{i,j} : (i,j)\in V)
\]
 be a subtuple of $(W_{1,1}, W_{1,2}, \ldots, W_{N,N})$.

\section{A Motivating Example} \label{motivatingExamples}
We now consider a two-node network that consists of a forward channel and a reverse channel, where the nodes are indexed by $1$ and $2$. Node $1$ and node~$2$ transmit information to each other through the channels as follows. In each time slot, node~$1$ transmits symbol $X_1\in \mathcal{X}_1$ to node~$2$ through the forward channel characterized by a conditional probability distribution $q_{Y_2|X_1}^{(1)}$, where $X_1$ and $q_{Y_2|X_1}^{(1)}$ together define $Y_2\in \mathcal{Y}_2$, the output of the forward channel. In the same time slot, node~2 receives~$Y_2$ and then transmits symbol~$X_2\in\mathcal{X}_2$ to node~$1$ through the reverse channel characterized by a conditional probability distribution $q_{Y_1|X_1, X_2, Y_2}^{(2)}$, where $(X_1, X_2, Y_2)$ and $q_{Y_1|X_1, X_2, Y_2}^{(2)}$ together define $Y_1\in \mathcal{Y}_1$, the output of the reverse channel.
Since node~$2$ receives~$Y_2$ before transmitting~$X_2$, $X_2$ can depend on $Y_2$. In other words, node~$2$ does not incur a delay and therefore the classical cut-set bound cannot be applied to this two-node network.

To facilitate discussion, we call this network the \textit{discrete memoryless channel (DMC) with noiseless reverse channel} if for all $x_1\in\mathcal{X}_1$, $x_2\in \mathcal{X}_2$, $y_1\in \mathcal{Y}_1$ and $y_2\in \mathcal{Y}_2$,
\[
q^{(2)}_{Y_1|X_1, X_2, Y_2}(y_1|x_1, x_2, y_2)=\begin{cases} 1 & \text{if $y_1 = x_2$,} \\ 0 &
\text{otherwise.}\end{cases}
\]
Note that the DMC with noiseless reverse channel reduces to the DMC with feedback \cite{CoverBook,Yeung08Book} if node~2 transmits in each time slot the symbol it receives in the same time slot.

\section{Generalized Discrete Memoryless Network} \label{sectionDefinition}
In this paper, we consider a general network that consists of $N$ nodes. Let \[\mathcal{I}=\{1, 2, \ldots, N\}\] be the index set of the nodes.
The $N$ terminals exchange information in $n$ time slots as follows.
Node~$i$ chooses message \[W_{i,j}\in \{1, 2, \ldots, M_{i,j}\}\] and sends $W_{i,j}$ to node~$j$ for each $(i, j)\in \mathcal{I}\times \mathcal{I}$. We assume that each message $W_{i,j}$ is uniformly distributed over $\{1, 2, \ldots, M_{i,j}\}$ and all the messages are independent. For each $k\in \{1, 2, \ldots, n\}$ and each $i\in \mathcal{I}$, node~$i$ transmits $X_{i,k} \in \mathcal{X}_i$ and receives $Y_{i,k} \in \mathcal{Y}_i$ in the $k^{\text{th}}$ time slot where $\mathcal{X}_i$ and $\mathcal{Y}_i$ are some alphabets that depend on~$i$.
After~$n$ time slots, node~$i$ declares~$\hat W_{j,i}$ to be the
transmitted~$W_{j,i}$ based on $W_{\{i\}\times \mathcal{I}}$ and $Y_i^n$ for each $(i, j)\in \mathcal{I}\times \mathcal{I}$. To simplify notation, let $M_{\mathcal{I}\times\mathcal{I}}$ denote the $N^2$-dimensional tuple $(M_{1,1}, M_{1,2}, \ldots, M_{N,N})$.
\medskip
\begin{Definition}\label{defOrderedPartition}
An $\alpha$-dimensional tuple $(\mathcal{S}_1, \mathcal{S}_2, \ldots \mathcal{S}_\alpha)$ consisting of subsets of~$\mathcal{I}$ is called an \textit{$\alpha$-partition} if
$
\cup_{h=1}^\alpha \mathcal{S}_h = \mathcal{I}
$
 and
$
 \mathcal{S}_i\cap \mathcal{S}_j = \emptyset
$
  for all $i\ne j$.
\end{Definition}
\medskip

For any $(\mathcal{S}_1, \mathcal{S}_2, \ldots \mathcal{S}_\alpha)$ consisting of subsets of~$\mathcal{I}$, we let
\[
\mathcal{S}^h=\cup_{i=1}^h \mathcal{S}_i
\]
for each $h\in\{1, 2, \ldots, \alpha\}$ to facilitate discussion.
Let $T\subseteq \mathcal{I}$ be any set.
 For any random tuple \[(X_{1}, X_{2}, \ldots, X_{N}) \in \mathcal{X}_1\times \mathcal{X}_2 \times \ldots \times \mathcal{X}_N,\] we let
 \[
 X_T=(X_{i} : i\in T)\] be a subtuple of $(X_{1}, X_{2}, \ldots, X_{N})$.
  In addition, we let $x_{\mathcal{S}^h}$ be the realization of $X_{\mathcal{S}^h}$ for each $h\in\{1, 2, \ldots, \alpha\}$. Similarly,
for any $k\in \{1, 2, \ldots, n\}$ and any random tuple \[(X_{1,k}, X_{2,k}, \ldots, X_{N, k}) \in \mathcal{X}_1\times \mathcal{X}_2 \times \ldots \times \mathcal{X}_N,\] we let
\[
X_{T,k}=(X_{i,k} : i\in T)
 \]
 be a subtuple of $(X_{1,k}, X_{2,k}, \ldots, X_{N, k}) $. In addition, we let $x_{\mathcal{S}^h, k}$ be the realization of $X_{\mathcal{S}^h,k}$ for each $h\in\{1, 2, \ldots, \alpha\}$. Under the classical model, the discrete memoryless network (DMN) is characterized by one channel specified by $q_{Y_\mathcal{I}|X_\mathcal{I}}$ and the elements in the random tuples $(X_{\mathcal{I},k},Y_{\mathcal{I},k})$ are generated in the order
 \[
 X_{\mathcal{I},k},Y_{\mathcal{I},k}
 \]
in the $k^{\text{th}}$ time slot for each $k\in\{1, 2, \ldots, n\}$ where the channel $q_{Y_\mathcal{I}|X_\mathcal{I}}$ is invoked to generate $Y_{\mathcal{I},k}$ from $X_{\mathcal{I},k}$. In contrast, under the generalized model to be defined below, two $\alpha$-partitions $(\mathcal{S}_1, \mathcal{S}_2, \ldots \mathcal{S}_\alpha)$ and $(\mathcal{G}_1, \mathcal{G}_2, \ldots \mathcal{G}_\alpha)$ are fixed in advance and the DMN is characterized by $\alpha$ channels specified by
\[
q_{Y_{\mathcal{G}_1}|X_{\mathcal{S}^1}}^{(1)}, q_{Y_{\mathcal{G}_2}|X_{\mathcal{S}^2}, Y_{\mathcal{G}^1}}^{(2)},\ldots,q_{Y_{\mathcal{G}_\alpha}|X_{\mathcal{S}^\alpha},Y_{\mathcal{G}^{\alpha-1}}}^{(\alpha)}.
\]
Under the generalized model, the elements in the random tuples $(X_{\mathcal{I},k},Y_{\mathcal{I},k})$ are generated in the order
\[
X_{\mathcal{S}_1,k}, Y_{\mathcal{G}_1,k}, X_{\mathcal{S}_2,k}, Y_{\mathcal{G}_2,k}, \ldots, X_{\mathcal{S}_\alpha,k}, Y_{\mathcal{G}_\alpha,k}
\]
in the $k^{\text{th}}$ time slot where the channel $q_{Y_{\mathcal{G}_h}|X_{\mathcal{S}^h}, Y_{\mathcal{G}^{h-1}}}^{(h)}$ is invoked to generate $Y_{\mathcal{G}_h,k}$ from $(X_{\mathcal{S}^h,k}, Y_{\mathcal{G}^{h-1},k})$ for each $h\in \{1, , \ldots, \alpha\}$.
\medskip
\begin{Definition} \label{defDiscreteNetwork}
The discrete network consists of $N$ finite input sets
$\mathcal{X}_1, \mathcal{X}_2, \ldots, \mathcal{X}_N$, $N$ finite output sets \linebreak $\mathcal{Y}_1, \mathcal{Y}_2, \ldots, \mathcal{Y}_N$ and $\alpha$ channels characterized by conditional distributions $q_{Y_{\mathcal{G}_1}|X_{\mathcal{S}^1}}^{(1)}, q_{Y_{\mathcal{G}_2}|X_{\mathcal{S}^2}, Y_{\mathcal{G}^1}}^{(2)},\ldots$, \linebreak $q_{Y_{\mathcal{G}_\alpha}|X_{\mathcal{S}^\alpha}, Y_{\mathcal{G}^{\alpha-1}}}^{(\alpha)}$, where $(\mathcal{S}_1, \mathcal{S}_2, \ldots \mathcal{S}_\alpha)$, denoted by $\boldsymbol{\mathcal{S}}$, and $(\mathcal{G}_1, \mathcal{G}_2, \ldots \mathcal{G}_\alpha)$, denoted by $\boldsymbol{\mathcal{G}}$, are two $\alpha$-partitions. We call $\boldsymbol{\mathcal{S}}$ and $\boldsymbol{\mathcal{G}}$ the \textit{input partition} and the \textit{output partition} of the network respectively. The discrete network is denoted by $(\mathcal{X}_\mathcal{I}, \mathcal{Y}_\mathcal{I}, \alpha, \boldsymbol{\mathcal{S}}, \boldsymbol{\mathcal{G}}, \boldsymbol q)$ where
  \[
  \boldsymbol q \triangleq (q^{(1)}, q^{(2)}, \ldots, q^{(\alpha)}).
  \]
\end{Definition}
\medskip
When we formally define a code on the discrete network later, we associate a tuple $B=(b_1, b_2, \ldots , b_N)$  called the \textit{delay profile} to the code and $b_i$ represents the amount of delay incurred by node~$i$ for the code. Under the classical model, $B$ consists of only positive numbers, meaning that the amount of delay incurred by each node is positive. In contrast, under our generalized model some elements of $B$ can take $0$ as long as no deadlock loop occurs. Therefore our model is a generalization of the classical model. A delay profile is formally defined as follows.
\medskip
\begin{Definition} \label{defDelayProfile}
A delay profile is an $N$-dimensional tuple $(b_1, b_2, \ldots , b_N)$ where $b_i \in \{0, 1\}$ for each $i\in \mathcal{I}$. The \textit{positive delay profile} is defined to be the $N$-dimensional all-one tuple denoted by $\mathbf{1}$.
\end{Definition}
\medskip
 The essence of the following definition is to characterize delay profiles which will not cause any deadlock loop for the transmissions for a given discrete network under the generalized model.
\medskip
\begin{Definition} \label{defFeasible}
Let $(\mathcal{X}_\mathcal{I}, \mathcal{Y}_\mathcal{I}, \alpha, \boldsymbol{\mathcal{S}}, \boldsymbol{\mathcal{G}}, \boldsymbol q)$ be a discrete network. For each $i\in \mathcal{I}$, let $h_i$ and $m_i$ be the two unique integers such that $i\in \mathcal{S}_{h_i}$ and $i\in\mathcal{G}_{m_i}$. Then, a delay profile $(b_1, b_2, \ldots, b_N)$ is said to be \textit{feasible for the network} if the following holds for each $i\in \mathcal{I}$: If $b_i=0$, then $h_i > m_i$.
\end{Definition}
\medskip
Under the classical model, the only delay profile considered is the positive delay profile $\mathbf{1}$, which is feasible for the network by Definition~\ref{defFeasible} (i.e., contains no deadlock loop). Under the generalized model, a delay profile may consist of zeros and a deadlock loop may occur if the delay profile consists of too many zeros. In the most extreme case where the delay profile is an all-zero tuple, every node will wait for the other nodes to transmit first before transmitting its own symbol, thus creating a deadlock. We are ready to define codes that use the network $n$ times in a deadlock-free manner as follows.
\medskip
\begin{Definition} \label{defCode}
Let $B\triangleq (b_1, b_2, \ldots , b_N)$ be a delay profile feasible for $(\mathcal{X}_\mathcal{I}, \mathcal{Y}_\mathcal{I}, \alpha, \boldsymbol{\mathcal{S}}, \boldsymbol{\mathcal{G}}, \boldsymbol q)$. A $(B, n, M_{\mathcal{I}\times \mathcal{I}})$-code, where $M_{\mathcal{I}\times \mathcal{I}}\triangleq (M_{1,1}, M_{1,2}, \ldots, M_{N,N})$ denotes the $N^2$-dimensional tuple of message alphabets, for $n$ uses of the network consists of the following:
\begin{enumerate}
\item A message set \[\mathcal{W}_{i,j}=\{1, 2, \ldots, M_{i,j}\}\] at node~$i$ for each $(i,j)\in \mathcal{I} \times \mathcal{I}$, where the message $W_{i,j}$ is uniform on $\mathcal{W}_{i,j}$. All the $N^2$ messages are independent.

\item An encoding function \[f_{i,k} : \mathcal{W}_{\{i\}\times\mathcal{I}} \times \mathcal{Y}_i^{k-b_i} \rightarrow \mathcal{X}_i\] for each $i\in \mathcal{I}$ and each $k\in\{1, 2, \ldots, n\}$, where $f_{i,k}$ is the encoding function at node~$i$ in the
$k^{\text{th}}$ time slot such that
\[
X_{i,k}=f_{i,k} (W_{\{i\}\times \mathcal{I}},
Y_i^{k-b_i}).
\]

\item A decoding function \[g_{i,j} : \mathcal{W}_{\{j\}\times\mathcal{I}} \times
\mathcal{Y}_j^{n} \rightarrow \mathcal{W}_{i,j}\] for each $(i, j) \in \mathcal{I}\times \mathcal{I}$, where $g_{i,j}$ is the decoding function for $W_{i,j}$ at node~$j$ such that
\[
\hat W_{i, j}=g_{i,j}(W_{\{j\}\times \mathcal{I}}, Y_j^{n}).
\]
\end{enumerate}
\end{Definition}
\medskip

Given a $(B, n, M_{\mathcal{I}\times\mathcal{I}})$-code, it follows from Definition~\ref{defCode} that for each $i\in\mathcal{I}$, node~$i$ incurs a delay if $b_i>0$, where $b_i$ is the amount of delay incurred by node~$i$. If $b_i=0$, node~$i$ incurs no delay, i.e., node~$i$ needs to receive $Y_{i,k}$ before encoding $X_{i,k}$ for each $k\in \{1, 2, \ldots, n\}$.
The feasibility condition of $B$ in Definition~\ref{defFeasible} ensures that the operations of any $(B, n, M_{\mathcal{I}\times\mathcal{I}})$-code are well-defined for the subsequently defined discrete memoryless network; the associated coding scheme is described after the network is defined.
\clearpage
\begin{Definition}\label{defMemoryless}
A discrete network $(\mathcal{X}_\mathcal{I}, \mathcal{Y}_\mathcal{I}, \alpha, \boldsymbol{\mathcal{S}}, \boldsymbol{\mathcal{G}}, \boldsymbol q)$, when used multiple times, is called a \textit{discrete memoryless network (DMN)} if the following holds for any $(B, n, M_{\mathcal{I}\times\mathcal{I}})$-code:

Let $U^{k-1}\triangleq (W_{\mathcal{I}\times \mathcal{I}}, X_{\mathcal{I}}^{k-1}, Y_{\mathcal{I}}^{k-1})$ be the collection of random variables that are generated before the $k^{\text{th}}$ time slot. Then, for each $k\in\{1, 2, \ldots, n\}$ and each $h\in \{1, 2, \ldots, \alpha\}$,
\begin{align}
& \Pr\{U^{k-1} = u^{k-1}, X_{\mathcal{S}^h,k} =x_{\mathcal{S}^h,k}, Y_{\mathcal{G}^{h},k}=y_{\mathcal{G}^{h},k} \}
 \notag\\
 & =  \Pr\{U^{k-1} = u^{k-1}, X_{\mathcal{S}^h,k} =x_{\mathcal{S}^h,k}, Y_{\mathcal{G}^{h-1},k}=y_{\mathcal{G}^{h-1},k} \} q_{Y_{\mathcal{G}_h}|X_{\mathcal{S}^h}, Y_{\mathcal{G}^{h-1}}}^{(h)}(y_{\mathcal{G}_{h},k}|x_{\mathcal{S}^h,k}, y_{\mathcal{G}^{h-1},k}) \label{memorylessStatement}
\end{align}
for all $u^{k-1}\in \mathcal{U}^{k-1}$, $x_{\mathcal{S}^h,k}\in \mathcal{X}_{\mathcal{S}^h}$ and $y_{\mathcal{G}^h,k}\in \mathcal{Y}_{\mathcal{G}^h}$.
\end{Definition}
\medskip

Following the notation in Definition~\ref{defMemoryless}, consider any $(B, n, M_{\mathcal{I}\times\mathcal{I}})$-code on the DMN. In the $k^{\text{th}}$ time slot, $X_{\mathcal{I},k}$ and $Y_{\mathcal{I},k}$ are generated in the order
\begin{equation}
X_{\mathcal{S}_1,k}, Y_{\mathcal{G}_1,k}, X_{\mathcal{S}_2,k}, Y_{\mathcal{G}_2,k}, \ldots, X_{\mathcal{S}_\alpha,k}, Y_{\mathcal{G}_\alpha,k} \label{orderExplanation}
\end{equation}
by transmitting on the channels in this order $q^{(1)},q^{(2)}, \ldots, q^{(\alpha)}$ using the $(B, n, M_{\mathcal{I}\times\mathcal{I}})$-code (as prescribed in Definition~\ref{defCode}). Specifically, $X_{\mathcal{S}^h, k}$, $Y_{\mathcal{G}^{h-1},k}$ and channel $q^{(h)}$ together define $Y_{\mathcal{G}_h,k}$ for each $h\in\{1, 2, \ldots, \alpha\}$. We will show in the following that the encoding of $X_{\mathcal{S}_h,k}$ before the transmission on $q^{(h)}$ and the generation of $Y_{\mathcal{G}_h,k}$ after the transmission on $q^{(h)}$ for each $h\in\{1, 2, \ldots, \alpha\}$ are well-defined.
Fix any $k\in\{1, 2, \ldots, n\}$ and $h\in\{1, 2, \ldots, \alpha\}$. Consider the following two cases for encoding $X_{i,k}$ for each $i\in \mathcal{S}_h$:\medskip\\
\textbf{Case \boldmath{$b_i>0$}:} Since $X_{i,k}$ is a function of $(W_{\{i\}\times \mathcal{I}}, Y_i^{k-b_i})$ and $Y_i^{k-b_i}$ has already been received by node~$i$ by the $k^{\text{th}}$ time slot, the encoding of $X_{i,k}$ at node~$i$ before the transmission on $q^{(h)}$ in the $k^{\text{th}}$ time slot is well-defined.\medskip \\
\textbf{Case \boldmath{$b_i=0$}:} Let $m$ be the unique integer such that
$i\in \mathcal{G}_m$.
 By the feasibility of~$B$, we have
 \begin{equation}
 h>m. \label{hStrictlyGreaterThanMTemp}
 \end{equation}
It follows from $i\in \mathcal{G}_m$ that $Y_{i,k}$ has already been received by node~$i$ before the transmission on $q^{(m+1)}$ in the $k^{\text{th}}$ time slot, which then implies from \eqref{hStrictlyGreaterThanMTemp} that $Y_{i,k}$ has already been received by node~$i$ before the transmission on $q^{(h)}$ in the $k^{\text{th}}$ time slot. Since $X_{i,k}$ is a function of $(W_{\{i\}\times \mathcal{I}}, Y_i^{k})$, $Y_{i,k}$ has already been received by node~$i$ before the transmission on $q^{(h)}$ in the $k^{\text{th}}$ time slot and $Y_i^{k-1}$ has already been received by node~$i$ by the $k^{\text{th}}$ time slot, it follows that the encoding of $X_{i,k}$ at node~$i$ before the transmission on $q^{(h)}$ in the $k^{\text{th}}$ time slot is well-defined. \medskip\\
Combining the two cases, the encoding of $X_{i,k}$ before the transmission on $q^{(h)}$ in the $k^{\text{th}}$ time slot for each $i\in \mathcal{S}_h$ is well-defined, which implies that the encoding of $X_{\mathcal{S}_h,k}$ before the transmission on $q^{(h)}$ in the $k^{\text{th}}$ time slot is well-defined.

In addition, the transmission on $q^{(h)}$ in the $k^{\text{th}}$ time slot only depends on $(X_{\mathcal{S}^{h},k}, Y_{\mathcal{G}^{h-1},k})$. Since the transmissions on $q^{(1)}, q^{(2)}, \ldots, q^{(h-1)}$ and the encoding of $X_{\mathcal{S}_1,k},X_{\mathcal{S}_2,k},\ldots, X_{\mathcal{S}_h,k}$ occur before the transmission on $q^{(h)}$ in the $k^{\text{th}}$ time slot, it follows that $Y_{\mathcal{G}^{h-1},k}$ and $X_{\mathcal{S}^h,k}$ have already been generated before the generation of $Y_{\mathcal{G}_h,k}$ according to \eqref{orderExplanation}, which implies that the generation of $Y_{\mathcal{G}_h,k}$ is well-defined.
\medskip
\begin{Example}\label{exampleTwoNode}
Consider a two-node DMN $(\mathcal{X}_{\{1,2\}}, \mathcal{Y}_{\{1,2\}}, 2, (\{1\},\{ 2\}), (\{2\}, \{1\}), (q^{(1)}, q^{(2)}))$ where all the alphabets are binary,
\begin{equation}
q^{(1)}_{Y_2|X_1}(y_2|x_1) =
\begin{cases}
1-\epsilon & \text{if $y_2 = x_1$,} \\
 \epsilon & \text{otherwise}
  \end{cases} \label{definitionP1}
\end{equation}
and
\begin{equation}
q^{(2)}_{Y_1|X_1, X_2, Y_2}(y_1|x_1, x_2, y_2) =
\begin{cases}
1 & \text{if $y_1 = x_2 + y_2$,} \\
 0 & \text{otherwise.}
  \end{cases}\label{definitionP2}
\end{equation}
Note that $q^{(1)}_{Y_2|X_1}$ is the conditional probability distribution for the binary symmetric channel (BSC). To facilitate discussion, we call this network the \textit{BSC with correlated feedback} which is illustrated in Figure~\ref{BSCFB}.
\begin{figure}[!t]
\center
\includegraphics[width=2.5 in, height=1.3 in, bb = 203 258 532 425, angle=0, clip=true]{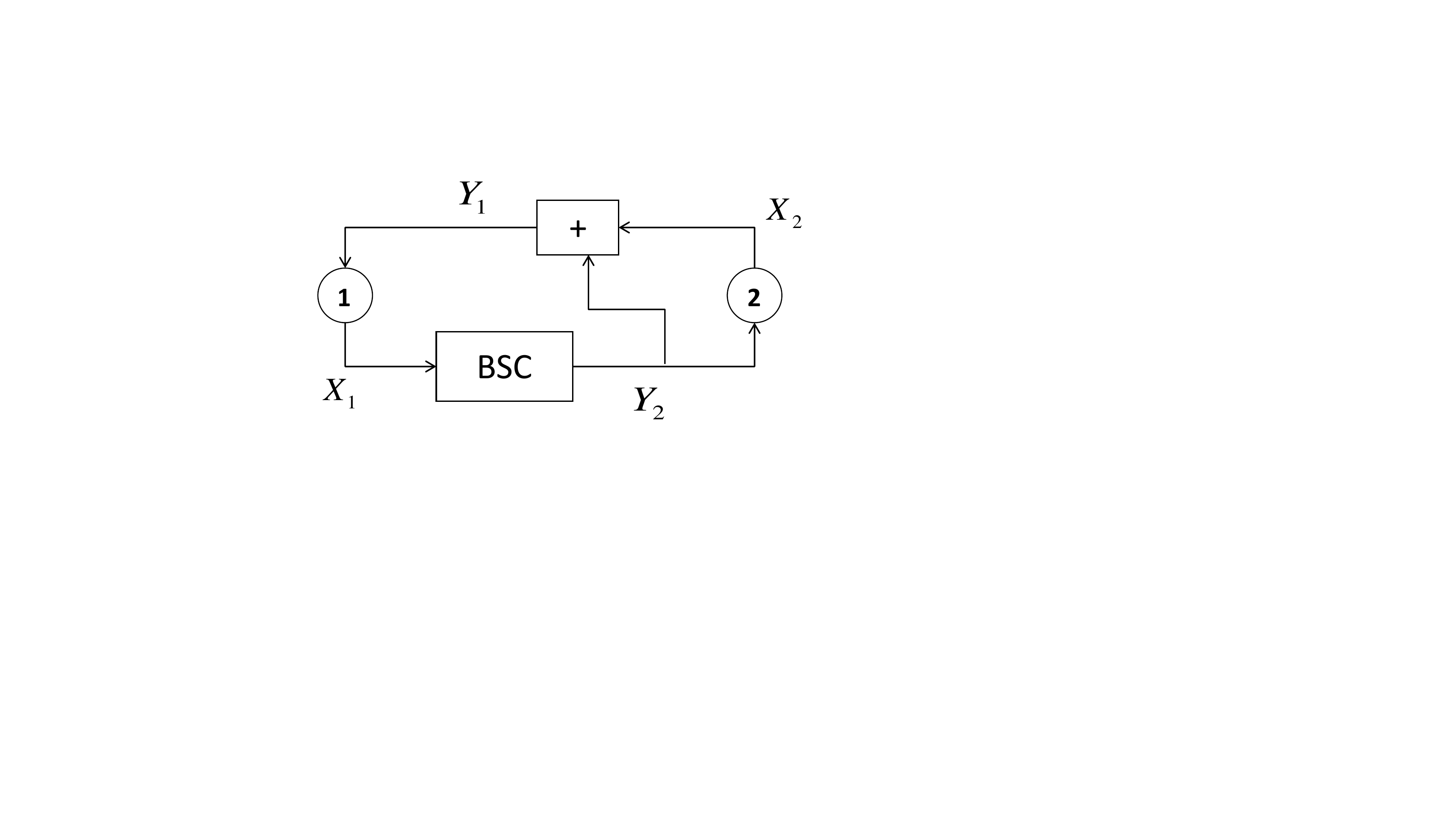}
\caption{BSC with correlated feedback.}
\label{BSCFB}
\end{figure}
For any $((1, 0), n, M_{\{1,2\}\times\{1,2\}})$-code on this network, $X_{\{1, 2\}, k}$ and $Y_{\{1, 2\}, k}$ are generated in the $k^{\text{th}}$ time slot in the order
\[
X_{1, k}, Y_{2, k}, X_{2, k}, Y_{1, k}
 \]
 for each $k\in\{1, 2, \ldots, n\}$. Note that node~$2$ incurs no delay, and can use $Y_{2,k}$ for encoding $X_{2,k}$ because $Y_{2, k}$ is generated before the generation of $X_{2, k}$. \hfill \small $\blacksquare$
\end{Example}
\medskip

In the classical model of the DMN, every node incurs a delay and the network is characterized by a single channel $q^{(1)}_{Y_{\mathcal{I}}|X_{\mathcal{I}}}$. Therefore, the classical DMN can be viewed as a generalized DMN with a single channel $q^{(1)}_{Y_{\mathcal{I}}|X_{\mathcal{I}}}$, and every code on the classical DMN can be viewed as some $(B, n, M_{\mathcal{I}\times \mathcal{I}})$-code on the generalized DMN with $B=\mathbf{1}$ (cf.\ Definitions~\ref{defDiscreteNetwork}, \ref{defFeasible}, \ref{defCode} and \ref{defMemoryless}).
\section{Capacity Region and Positive-Delay Region and Their Cut-Set Bounds} \label{sectionCapacityRegions}
 Besides the capacity region of the DMN, we are also interested in the positive-delay region -- the set of achievable rate tuples under the constraint that every node incurs a delay. We formally define the capacity region and the positive-delay region through the following three intuitive definitions.
\medskip
\begin{Definition} \label{cutseterrorProbability}
For a $(B, n, M_{\mathcal{I}\times\mathcal{I}})$-code on the DMN, the average probability of decoding error of~$W_{i,j}$ is defined as
$
P_{i,j}^n = \Pr\{ g_{i,j}(W_{\{j\}\times \mathcal{I}}, Y_j^n) \ne W_{i,j} \}
$ for each $(i,j)\in \mathcal{I}\times\mathcal{I}\,$.
\end{Definition}
\medskip
\begin{Definition} \label{cutsetachievable rate}
A rate tuple $(R_{1,1}, R_{1,2}, \ldots, R_{N,N})$, denoted by $R_{\mathcal{I}\times \mathcal{I}}$, is \textit{$B$-achievable} for the DMN if there exists a sequence of $(B, n, M_{\mathcal{I}\times\mathcal{I}})$-codes with
$
 \lim\limits_{n\rightarrow \infty} \frac{\log  M_{i,j}}{n} \ge R_{i,j}
$
 such that
$
 \lim\limits_{n\rightarrow \infty} P_{i,j}^n = 0
$
for each $(i,j)\in \mathcal{I}\times\mathcal{I}$. The tuple $R_{\mathcal{I}\times \mathcal{I}}$ is said to be \textit{achievable} for the DMN if it is $B$-achievable for some delay profile~$B$.
\end{Definition}
\medskip

Without loss of generality, we assume that $M_{i,i}=1$ and $R_{i,i}=0$ for all $i\in \mathcal{I}$ in the rest of this paper.
\medskip
\begin{Definition}\label{cutsetcapacity region}
The \textit{$B$-achievable rate region}, denoted by $\mathcal{R}_B$, of the DMN is the closure of the set consisting of every $B$-achievable rate tuple $R_{\mathcal{I}\times \mathcal{I}}$ with $R_{i,i}=0$ for all $i\in \mathcal{I}$. We call
\[
\mathcal{C} \triangleq \bigcup_{B: \text{ $B$ is feasible for the DMN}} \mathcal{R}_B
\]
the \textit{capacity region}, and we call
$
\mathcal{R}_{\mathbf{1}}
$
the \textit{positive-delay region}.
\end{Definition}
\medskip
The following two theorems are the main results of this paper.
\medskip
\begin{Theorem}\label{cutsetThmCapacity}
 Let
 \begin{align}
\mathcal{R}^{\text{out}}& \triangleq \bigcup_{\substack{p_{X_{\mathcal{I}}, Y_{\mathcal{I}}}:p_{X_{\mathcal{I}}, Y_{\mathcal{I}}}=\\ \prod_{h=1}^\alpha (p_{X_{\mathcal{S}_h}|X_{\mathcal{S}^{h-1}}, Y_{\mathcal{G}^{h-1}}} q^{(h)}_{Y_{\mathcal{G}_h} | X_{\mathcal{S}^h},  Y_{\mathcal{G}^{h-1}}}) }} \bigcap_{T\subseteq \mathcal{I}} \notag\\
& \quad \qquad \left\{ R_{\mathcal{I}\times \mathcal{I}}\left| \: \parbox[c]{4.45 in}{$ \sum\limits_{ (i,j)\in T\times T^c}  R_{i,j}
 \le \sum\limits_{h=1}^\alpha  I_{p_{X_{\mathcal{I}}, Y_{\mathcal{I}}}}(X_{T\cap \mathcal{S}^h}, Y_{T\cap \mathcal{G}^{h-1}} ;  Y_{T^c\cap \mathcal{G}_h}| X_{T^c \cap \mathcal{S}^h},  Y_{T^c\cap \mathcal{G}^{h-1}})
 $} \right.\right\}. \label{Rout}
\end{align}
Then,
\begin{equation}
\mathcal{C}\subseteq\mathcal{R}^{\text{out}}. \label{thmResultCapacity}
\end{equation}
\end{Theorem}
\medskip
\begin{Theorem}\label{cutsetThmPositive}
Let
 \begin{align}
\mathcal{R}_{\mathbf{1}}^{\text{out}}& \triangleq \bigcup_{\substack{p_{X_{\mathcal{I}}, Y_{\mathcal{I}}}:p_{X_{\mathcal{I}}, Y_{\mathcal{I}}}=\\ p_{X_{\mathcal{I}}} \prod_{h=1}^\alpha q^{(h)}_{Y_{\mathcal{G}_h} | X_{\mathcal{S}^h},  Y_{\mathcal{G}^{h-1}}}}} \bigcap_{T\subseteq \mathcal{I}} \left\{ R_{\mathcal{I}\times \mathcal{I}}\left| \: \parbox[c]{2.5 in}{$ \sum\limits_{ (i,j)\in T\times T^c}  R_{i,j}
 \le  I_{p_{X_{\mathcal{I}}, Y_{\mathcal{I}}}}(X_T;  Y_{T^c}| X_{T^c})
 $} \right.\right\}. \label{RoutPositive}
\end{align}
Then,
\begin{equation}
\mathcal{R}_\mathbf{1}\subseteq\mathcal{R}_\mathbf{1}^{\text{out}}. \label{thmResultPositive}
\end{equation}
\end{Theorem}
\medskip
\begin{Remark}
Theorems~\ref{cutsetThmCapacity} and~\ref{cutsetThmPositive} state the cut-set outer bounds for $\mathcal{C}$ and $\mathcal{R}_1$ respectively. More specifically, $\mathcal{R}^{\text{out}}$ in \eqref{Rout} and $\mathcal{R}_{\mathbf{1}}^{\text{out}}$ in \eqref{RoutPositive} are the cut-set bounds for $\mathcal{C}$ and $\mathcal{R}_1$ respectively.
\end{Remark}
\medskip
\begin{Remark}
Every DMN defined under the classical model can be viewed as a generalized DMN characterized by some single channel denoted by $q^{(1)}_{Y_{\mathcal{I}}|X_{\mathcal{I}}}$. For such a DMN, it follows from Definition~\ref{defFeasible} that the only feasible delay profile is $\mathbf{1}$ and from Definition~\ref{defOrderedPartition} that $\mathcal{S}_1=\mathcal{G}_1=\mathcal{I}$, which then imply respectively that $\mathcal{C}=\mathcal{R}_\mathbf{1}$ and $\mathcal{R}^{\text{out}}=\mathcal{R}_\mathbf{1}^{\text{out}}$. Consequently, Theorem~\ref{cutsetThmCapacity} and Theorem~\ref{cutsetThmPositive} are identical for the classical DMN and they yield the same cut-set bound
\[
\mathcal{R}^{\text{out}}=\mathcal{R}_\mathbf{1}^{\text{out}}=\bigcup_{p_{X_{\mathcal{I}}}}\: \bigcap_{T\subseteq \mathcal{I}}  \left\{ R_{\mathcal{I}\times \mathcal{I}}\left| \: \parbox[c]{2.65 in}{$ \sum\limits_{ (i,j)\in T\times T^c}  R_{i,j}
 \le I_{p_{X_{\mathcal{I}}} q^{(1)}_{Y_{\mathcal{I}} | X_{\mathcal{I}}}}(X_{T};  Y_{T^c}| X_{T^c})
 $} \right.\right\},
\]
 which coincides with the classical cut-set bound in \cite[Chapter 15]{CoverBook}. If we consider the DMN under the multimessage multicast scenario as described in \cite[Section 18.4.2]{elgamalBook} where each node has only a single message (instead of~$N$ messages considered in Theorems~\ref{cutsetThmCapacity} and~\ref{cutsetThmPositive}) to transmit, then we can follow similar techniques for proving Theorem~\ref{cutsetThmCapacity} to obtain the cut-set bound for the generalized multimessage multicast DMN, whose statement and proof are contained in \cite[Theorem 2]{fong14GeneralizedMMN}. In particular, if the generalized multimessage multicast DMN is characterized by only one channel, then the cut-set bound obtained for the multicast DMN coincides with the classical cut-set bound stated in \cite[Section 18.4.2]{elgamalBook}.
 \end{Remark}
 \medskip
\begin{Remark}
Using Theorem~\ref{cutsetThmCapacity}, we obtain
\[
R_{1,2} \le \max_{p_{X_1}}I_{p_{X_1}q_{Y_2|X_1}^{(1)}}(X_1;Y_2)
 \]
  for the DMC with noiseless reverse channel described in Section~\ref{motivatingExamples}, which implies the well-known result that the presence of feedback does not increase the capacity for the DMC $q_{Y_2|X_1}^{(1)}$ \cite{Sha56}.
  \end{Remark}
\medskip
\begin{Remark}
We will use Theorem~\ref{cutsetThmCapacity} to prove Theorem~\ref{cutsetThmPositive}. To this end, we will prove a folklore theorem which states that under the constraint that every node incurs a delay, the generalized DMN characterized by $\alpha$ channels $q^{(1)}, q^{(2)}, \ldots, q^{(\alpha)}$ is equivalent to the classical DMN characterized by a single channel $q^{(1)} q^{(2)} \ldots q^{(\alpha)}$. This will be shown in Section~\ref{sectionOuterBound*}.
\end{Remark}

\section{Proof of Cut-Set Bound on Capacity Region}\label{sectionOuterBound}
A complete proof of Theorem~\ref{cutsetThmCapacity} is presented in this section. The following lemma and two propositions are preparations for the proof of Theorem~\ref{cutsetThmCapacity}.
The following proposition characterizes an important property of Markov chains.
\medskip
\begin{Proposition} \label{propositionMCsimplification}
Suppose there exist two probability distributions $r_{X,Y}$ and $q_{Z|Y}$ such that
\begin{equation}
p_{X, Y, Z}(x,y,z) = r_{X,Y}(x,y)q_{Z|Y}(z|y) \label{statement1CorollaryMC}
\end{equation}
for all $x$, $y$ and $z$ whenever $p_Y(y)>0$. Then
\begin{equation}
(X\rightarrow Y\rightarrow Z)_{p_{X,Y,Z}} \label{statement3CorollaryMC}
\end{equation}
forms a Markov chain. In addition,
\begin{equation}
p_{Z|Y}=q_{Z|Y} \label{statement2CorollaryMC}.
\end{equation}
\end{Proposition}
\begin{IEEEproof}The proof of \eqref{statement3CorollaryMC} is contained in \cite[Proposition~2.5]{Yeung08Book}. It remains to show \eqref{statement2CorollaryMC}. Summing $x$ and then $z$ on both sides of \eqref{statement1CorollaryMC}, we have $p_{Y,Z}(y,z)=r_Y(y)q_{Z|Y}(z|y)$ and $p_Y(y) = r_Y(y)$ for all $x$, $y$ and $z$ whenever $p_Y(y)>0$, which implies \eqref{statement2CorollaryMC}.
\end{IEEEproof}
\medskip
 The following lemma is a direct consequence of the memoryless property of the DMN stated in Definition~\ref{defMemoryless}.
 \medskip
\begin{Lemma}\label{cutsetMCLemma}
Let $(\mathcal{X}_\mathcal{I}, \mathcal{Y}_\mathcal{I}, \alpha, \boldsymbol{\mathcal{S}}, \boldsymbol{\mathcal{G}}, \boldsymbol{q})$ be a DMN. Fix any $(B, n, M_{\mathcal{I}\times\mathcal{I}})$-code on the DMN and let $p_{X_\mathcal{I}, Y_\mathcal{I}}$ denote the distribution induced by the code. Then, for each $k\in\{1, 2, \ldots, n\}$ and each $h\in\{1, 2, \ldots, \alpha\}$,
\begin{equation}
p_{Y_{\mathcal{G}_h,k} |X_{\mathcal{S}^h,k}, Y_{\mathcal{G}^{h-1},k}}(y_{\mathcal{G}_h,k} |x_{\mathcal{S}^h,k}, y_{\mathcal{G}^{h-1},k})
 = q_{Y_{\mathcal{G}_h} |X_{\mathcal{S}^h}, Y_{\mathcal{G}^{h-1}}}^{(h)}(y_{\mathcal{G}_h,k} |x_{\mathcal{S}^h,k}, y_{\mathcal{G}^{h-1},k}) \label{thirdStatement}
\end{equation}
for all $x_{\mathcal{S}^h,k}$ and $y_{\mathcal{S}^h,k}$.
\end{Lemma}
\begin{IEEEproof}
Let $U^{k-1}=(W_{\mathcal{I}\times \mathcal{I}}, X_{\mathcal{I}}^{k-1}, Y_{\mathcal{I}}^{k-1})$ be the collection of random variables that are generated before the $k^{\text{th}}$ time slot for the $(B, n, M_{\mathcal{I}\times\mathcal{I}})$-code. It follows from Definition~\ref{defMemoryless} that for each $k\in\{1, 2, \ldots, n\}$ and each $h \in \{1, 2, \ldots, \alpha\}$,
\begin{align}
& p_{U^{k-1}, X_{\mathcal{S}^h,k}, Y_{\mathcal{G}^{h},k}}(u^{k-1}, x_{\mathcal{S}^h,k}, y_{\mathcal{G}^{h},k})\notag\\
& = p_{U^{k-1}, X_{\mathcal{S}^h,k}, Y_{\mathcal{G}^{h-1},k} }(u^{k-1}, x_{\mathcal{S}^h,k}, y_{\mathcal{G}^{h-1},k} )q_{Y_{\mathcal{G}_h,k} | X_{\mathcal{S}^h,k}, Y_{\mathcal{G}^{h-1},k}}^{(h)}(y_{\mathcal{G}_h,k} | x_{\mathcal{S}^h,k}, y_{\mathcal{G}^{h-1},k}) \label{lemmaTempStatement}
\end{align}
for all $u^{k-1}$, $x_{\mathcal{S}^h,k}$ and $y_{\mathcal{G}^{h-1},k}$.
Equation \eqref{thirdStatement} then follows from applying Proposition~\ref{propositionMCsimplification} to \eqref{lemmaTempStatement}.
\end{IEEEproof}
\medskip
The following proposition is a consequence of the definition of the $(B, n, M_{\mathcal{I}\times\mathcal{I}})$-code in Definition~\ref{defCode}.
\begin{Proposition}\label{propositionXFunctionOfY}
Fix any $(B, n, M_{\mathcal{I}\times\mathcal{I}})$-code on the DMN $(\mathcal{X}_\mathcal{I}, \mathcal{Y}_\mathcal{I}, \alpha, \boldsymbol{\mathcal{S}}, \boldsymbol{\mathcal{G}}, \boldsymbol{q})$ and fix an $h\in\{1, 2, \ldots, \alpha\}$. Then, for each $i \in \mathcal{S}_h$, $X_{i,k}$ is a function of $(W_{\{i\}\times \mathcal{I}}, Y_i^{k-1}, Y_{\{i\} \cap \mathcal{G}^{h-1},k})
$
 for each $k\in\{1, 2, \ldots, n\}$.
\end{Proposition}
\begin{IEEEproof}
Let $B=(b_1, b_2, \ldots , b_N)$. Fix an $i \in \mathcal{S}_h$. By Definition~\ref{defCode}, $X_{i,k}$ is a function of $(W_{\{i\}\times \mathcal{I}}, Y_i^{k-b_i})$ for each $k\in\{1, 2, \ldots, n\}$. Consider the following two cases: \smallskip \\
 \textbf{Case \boldmath{$b_i = 1$}:} Since $X_{i,k}$ is a function of $(W_{\{i\}\times \mathcal{I}}, Y_i^{k-b_i})$ and $b_i=1$, $X_{i,k}$ is a function of
\[
 (W_{\{i\}\times \mathcal{I}}, Y_i^{k-1}, Y_{\{i\} \cap \mathcal{G}^{h-1},k}).
\]
 \textbf{Case \boldmath{$b_i = 0$}:} Let $m$ be the unique integer such that $i\in\mathcal{G}_m$. Since $i\in\mathcal{S}_h$ and $B$ is feasible for the network (cf.\ Definition~\ref{defCode}), it follows from Definition~\ref{defFeasible} that
\begin{equation}
h>m. \label{hGreaterThanm}
\end{equation}
Since $i\in\mathcal{G}_m$ and $X_{i,k}$ is a function of $(W_{\{i\}\times \mathcal{I}}, Y_i^{k})$, $X_{i,k}$ is a function of
$
(W_{\{i\}\times \mathcal{I}}, Y_i^{k-1}, Y_{\{i\} \cap \mathcal{G}_m,k}),
$
 which implies from \eqref{hGreaterThanm} that $X_{i,k}$ is a function of
$
 (W_{\{i\}\times \mathcal{I}}, Y_i^{k-1}, Y_{\{i\} \cap \mathcal{G}^{h-1},k})$.
\end{IEEEproof}
\medskip
Equipped with Proposition~\ref{propositionMCsimplification}, Lemma~\ref{cutsetMCLemma} and Proposition~\ref{propositionXFunctionOfY}, we are now ready to prove Theorem~\ref{cutsetThmCapacity}.
\medskip
\begin{IEEEproof}[\textbf{Proof of Theorem~\ref{cutsetThmCapacity}}]
Suppose $R_{\mathcal{I}\times \mathcal{I}}$ is in $\mathcal{C}$. By Definitions~\ref{cutsetachievable rate} and~\ref{cutsetcapacity
region}, there exists a sequence of $(B, n, M_{\mathcal{I}\times\mathcal{I}})$-codes such that
\begin{equation}
 \lim_{n\rightarrow \infty} \frac{\log  M_{i,j}}{n} \ge R_{i,j} \label{thmTempEq1}
\end{equation}
and
\begin{equation}
 \lim_{n\rightarrow \infty} P_{i,j}^n = 0 \label{thmTempEq2}
\end{equation}
for each $(i,j)\in \mathcal{I}\times\mathcal{I}$.
Fix~$n$ and the corresponding $(B, n, M_{\mathcal{I}\times\mathcal{I}})$-code, and let $p_{W_{\mathcal{I}\times \mathcal{I}},X_\mathcal{I}^n, Y_\mathcal{I}^n, \hat W_{\mathcal{I}\times \mathcal{I}}}$ be the probability distribution induced by the code. Fix any $T\subseteq \mathcal{I}$.
Since the $N^2$ messages $W_{1,1}, W_{1,2}, \ldots, W_{N,N}$ are independent, we have
\begin{align}
&\!\!  \sum_{(i,j)\in T \times T^c} \log  M_{i,j}\notag \\
& \! = H_{p_{W_{\mathcal{I}\times \mathcal{I}}}}(W_{T\times T^c}|W_{(T\times T^c)^c})\notag \\
&\!= I_{p_{W_{\mathcal{I}\times \mathcal{I}},Y_{T^c}^n}}(W_{T\times T^c}; Y_{T^c}^n|W_{(T\times T^c)^c})\! + H_{p_{W_{\mathcal{I}\times \mathcal{I}},Y_{T^c}^n}}(W_{T\times T^c}|Y_{T^c}^n,W_{(T\times T^c)^c}) \notag\\
&\!\le I_{p_{W_{\mathcal{I}\times \mathcal{I}},Y_{T^c}^n}}(W_{T\times T^c}; Y_{T^c}^n|W_{(T\times T^c)^c})\! + H_{p_{W_{\mathcal{I}\times \mathcal{I}},Y_{T^c}^n}}(W_{T\times T^c}|Y_{T^c}^n,W_{T^c \times \mathcal{I}}) \notag\\
&\!\le I_{p_{W_{\mathcal{I}\times \mathcal{I}},Y_{T^c}^n}}(W_{T\times T^c}; Y_{T^c}^n|W_{(T\times T^c)^c}) +
\!\!\!\! \sum_{(i,j)\in T\times T^c} \!\!\!\!\! H_{p_{W_{\mathcal{I}\times \mathcal{I}},Y_{T^c}^n}}(W_{i,j}|Y_j^n, W_{\{j\} \times \mathcal{I}})\notag \\
 &\!\le   I_{p_{W_{\mathcal{I}\times \mathcal{I}},Y_{T^c}^n}}(W_{T\times T^c}; \! Y_{T^c}^n|W_{(T\times T^c)^c})\! +
\!\!\!\!\!\!  \sum_{(i,j)\in T\times T^c} \! \!\!\!\! (1 \! + P_{i,j}^n \log   M_{i,j})
\label{cutseteqnSet1}
\end{align}
where the last inequality follows from Fano's inequality (cf.\ Definition~\ref{cutseterrorProbability}).
Following \eqref{cutseteqnSet1} and omitting the subscripts for the entropy and mutual information terms, we consider
\begin{align}
 & I(W_{T\times T^c}; Y_{T^c}^n|W_{(T\times T^c)^c})\notag \\
&\quad=  \sum_{k=1}^n  (H(Y_{T^c,k}|W_{(T\times T^c)^c}, Y_{T^c}^{k-1})  -H(Y_{T^c,k}|W_{\mathcal{I}\times \mathcal{I}} ,Y_{T^c}^{k-1})) \notag \\
&\quad \stackrel{\text{(a)}}{=} \sum_{k=1}^n  (H(Y_{T^c \cap (\cup_{h=1}^\alpha \mathcal{G}_h),k}|W_{(T\times T^c)^c}, Y_{T^c}^{k-1}) -H(Y_{T^c \cap (\cup_{h=1}^\alpha \mathcal{G}_h),k}|W_{\mathcal{I}\times \mathcal{I}} ,Y_{T^c}^{k-1})) \notag \\
&\quad \le \sum_{k=1}^n \sum_{h=1}^\alpha (H(Y_{T^c\cap \mathcal{G}_h ,k}|W_{T^c \times \mathcal{I}}, Y_{T^c}^{k-1}, Y_{T^c\cap \mathcal{G}^{h-1} ,k})  -H(Y_{T^c\cap \mathcal{G}_h,k}|W_{\mathcal{I}\times \mathcal{I}}, Y_{T^c}^{k-1}, Y_{T^c \cap \mathcal{G}^{h-1},k} ))  \label{cutsetstatement8}
\end{align}
where (a) follows from the fact that $\cup_{h=1}^\alpha \mathcal{G}_h = \mathcal{I}\,.$ Following~\eqref{cutsetstatement8}, we obtain
\begin{align}
& H(Y_{T^c\cap \mathcal{G}_h ,k}|W_{T^c \times \mathcal{I}}, Y_{T^c}^{k-1}, Y_{T^c\cap \mathcal{G}^{h-1} ,k})\notag \\
&\, \stackrel{\text{(a)}}{=} H(Y_{T^c\cap \mathcal{G}_h ,k}|W_{T^c \times \mathcal{I}}, Y_{T^c}^{k-1}, Y_{T^c \cap \mathcal{G}^{h-1},k}, X_{T^c \cap \mathcal{S}^h,k} )\notag \\
&\, \le  H(Y_{T^c\cap \mathcal{G}_h ,k}| X_{T^c \cap \mathcal{S}^h,k},  Y_{T^c \cap \mathcal{G}^{h-1},k})\label{cutsetstatement88}
\end{align}
and
\begin{align}
& H(Y_{T^c\cap \mathcal{G}_h,k}|W_{\mathcal{I}\times \mathcal{I}}, Y_{T^c}^{k-1}, Y_{T^c \cap \mathcal{G}^{h-1},k} )\notag \\
&\quad\ge H(Y_{T^c\cap \mathcal{G}_h,k}|W_{\mathcal{I}\times \mathcal{I}}, X_{\mathcal{I}}^{k-1} ,X_{\mathcal{S}^{h},k} ,Y_{\mathcal{I}}^{k-1}, Y_{\mathcal{G}^{h-1},k})\notag \\
&\quad \stackrel{\text{(b)}}{=} H(Y_{T^c\cap \mathcal{G}_h,k}|X_{\mathcal{S}^{h},k}, Y_{\mathcal{G}^{h-1},k}) \label{cutsetstatement9}
\end{align}
for each $h\in\{1, 2, \ldots, \alpha\}$, where
 \begin{enumerate}
 \item[(a)] follows from Proposition \ref{propositionXFunctionOfY} that for each $\ell \in\{1, 2, \ldots, h\}$, $X_{T^c \cap \mathcal{S}_\ell,k} $ is a function of
     \[
     (W_{(T^c \cap \mathcal{S}_\ell) \times \mathcal{I}}, Y_{T^c \cap \mathcal{S}_\ell}^{k-1},  Y_{T^c \cap \mathcal{S}_\ell \cap \mathcal{G}^{\ell-1},k}).
     \]
\item[(b)] follows from Definition~\ref{defMemoryless} that for each $k\in\{1, 2, \ldots, n\}$ and each $h\in\{1, 2, \ldots, \alpha\}$,
\begin{equation*}
\left((W_{\mathcal{I}\times \mathcal{I}}, X_{\mathcal{I}}^{k-1}, Y_{\mathcal{I}}^{k-1})  \rightarrow (X_{\mathcal{S}^h,k},Y_{\mathcal{G}^{h-1},k}) \rightarrow Y_{\mathcal{G}_h,k}\right)_p
\end{equation*}
forms a Markov Chain.
\end{enumerate}
Define
\[
\tilde p_{Q_n}(k) = 1/n
\]
for each $k\in\{1, 2, \ldots, n\}$ where $Q_n$ is a timesharing random variable uniformly distributed
on $\{1, 2, \ldots, n\}$.
Construct
$\tilde p_{Q_n, X_{\mathcal{I}, Q_n}, Y_{\mathcal{I},Q_n}}$
such that
\begin{equation}
\tilde p_{Q_n, X_{\mathcal{I}, Q_n}, Y_{\mathcal{I},Q_n}}(k, x_\mathcal{I}, y_\mathcal{I}) \triangleq \tilde p_{Q_n}(k)p_{X_{\mathcal{I},k},Y_{\mathcal{I},k}}(x_\mathcal{I},y_\mathcal{I}) \label{defTildeP}
\end{equation}
for all $k\in\{1, 2, \ldots, n\}$, all $x_\mathcal{I}\in \mathcal{X}_\mathcal{I}$ and all $y_\mathcal{I}\in \mathcal{Y}_\mathcal{I}$ (recall that $p$ refers to the distribution induced by the code).
  Then, for any $\mathcal{A}, \mathcal{B} \subseteq \mathcal{I}$, it follows from \eqref{defTildeP} that
\begin{align}
\tilde p_{X_{\mathcal{A},Q_n}, Y_{\mathcal{B},Q_n}|Q_n}(x_\mathcal{A}, y_\mathcal{B} |k)
&  =p_{X_{\mathcal{A},k}, Y_{\mathcal{B},k}}(x_\mathcal{A}, y_\mathcal{B})
\label{cutsetMCQ_n}
\end{align}
for all $k\in\{1, 2, \ldots, n\}$, all $x_\mathcal{A}\in \mathcal{X}_\mathcal{A}$ and all $y_\mathcal{B}\in \mathcal{Y}_\mathcal{B}$,
which implies that
\begin{equation}
H_{\tilde p_{Q_n,X_{\mathcal{A},Q_n}, Y_{\mathcal{B},Q_n}}}(X_{\mathcal{A}, Q_n}, Y_{\mathcal{B}, Q_n}|Q_n = k) = H_{p_{X_{\mathcal{A},k}, Y_{\mathcal{B},k}}}(X_{\mathcal{A}, k}, Y_{\mathcal{B}, k}).\label{cutsetMCQ_n*}
\end{equation}
In addition,
\begin{align}
& \! \tilde p_{Q_n, X_{\mathcal{S}^h,Q_n}, Y_{\mathcal{G}^h,Q_n}}(k, x_{\mathcal{S}^h}, y_{\mathcal{G}^h})\notag\\
&  \!\! = \tilde p_{Q_n}(k)\tilde p_{X_{\mathcal{S}^h,Q_n}, Y_{\mathcal{G}^h,Q_n}|Q_n}(
x_{\mathcal{S}^h}, y_{\mathcal{G}^h} |k)  \notag
\\
& \!\!\stackrel{\eqref{cutsetMCQ_n}}{=} \tilde p_{Q_n}(k)p_{X_{\mathcal{S}^h,k}, Y_{\mathcal{G}^h,k}}(x_{\mathcal{S}^h}, y_{\mathcal{G}^h})\notag \\
& \!\! \stackrel{\text{(a)}}{=} \tilde p_{Q_n} (k)p_{X_{\mathcal{S}^h,k}, Y_{\mathcal{G}^{h-1},k}}(x_{\mathcal{S}^h}, y_{\mathcal{G}^{h-1}})q_{Y_{\mathcal{G}_h}|X_{\mathcal{S}^h}, Y_{\mathcal{G}^{h-1}}}^{(h)}(y_{\mathcal{G}_h}|x_{\mathcal{S}^h}, y_{\mathcal{G}^{h-1}}) \label{cutsetstatement12}
\end{align}
for each $h\in\{1, 2, \ldots, \alpha\}$  where (a) follows from Lemma~\ref{cutsetMCLemma}. It then follows from (\ref{cutsetstatement12}) and Proposition~\ref{propositionMCsimplification} that
\begin{equation}
\left(Q_n \rightarrow (X_{\mathcal{S}^h,Q_n},Y_{\mathcal{G}^{h-1}, Q_n}) \rightarrow Y_{\mathcal{G}_h,Q_n} \right)_{\tilde p} \label{MCTimeVariable}
\end{equation}
forms a Markov Chain.
Following \eqref{cutsetstatement88} and \eqref{cutsetstatement9}, we consider
\begin{align}
& \frac{1}{n}\sum_{k=1}^n \sum_{h=1}^\alpha H_{p_{X_{\mathcal{I},k}, Y_{\mathcal{I},k}}}(Y_{T^c\cap \mathcal{G}_h ,k}| X_{T^c \cap \mathcal{S}^h,k}, Y_{T^c \cap \mathcal{G}^{h-1},k}) \notag \\
 & \stackrel{\eqref{cutsetMCQ_n*}}{=}  \sum_{h=1}^\alpha \sum_{k=1}^n \! \frac{1}{n} H_{\tilde p_{Q_n, X_{\mathcal{I},Q_n}, Y_{\mathcal{I},Q_n}}}(Y_{T^c\cap \mathcal{G}_h ,Q_n}| X_{T^c \cap \mathcal{S}^h \! , Q_n}, Y_{T^c \cap \mathcal{G}^{h-1},Q_n}, Q_n \!\!= \!k)\notag \\
 & = \sum_{h=1}^\alpha H_{\tilde p_{Q_n, X_{\mathcal{I},Q_n}, Y_{\mathcal{I},Q_n}}}(Y_{T^c\cap \mathcal{G}_h ,Q_n}| X_{T^c \cap \mathcal{S}^h,Q_n}, Y_{T^c \cap \mathcal{G}^{h-1},Q_n}, Q_n)\notag \\
& \le \sum_{h=1}^\alpha H_{\tilde p_{X_{\mathcal{I},Q_n}, Y_{\mathcal{I},Q_n}}}(Y_{T^c\cap \mathcal{G}_h ,Q_n}| X_{T^c \cap \mathcal{S}^h, Q_n},Y_{T^c \cap \mathcal{G}^{h-1},Q_n}) \label{cutseteqnSet2}
\end{align}
and
\begin{align}
& \frac{1}{n}\sum_{k=1}^n \sum_{h=1}^\alpha H_{p_{X_{\mathcal{I},k}, Y_{\mathcal{I},k}}}(Y_{T^c\cap \mathcal{G}_h,k}|X_{\mathcal{S}^{h},k},Y_{\mathcal{G}^{h-1},k}) \notag \\
& \quad \stackrel{\eqref{cutsetMCQ_n*}}{=} \sum_{h=1}^\alpha \sum_{k=1}^n \frac{1}{n} H_{\tilde p_{Q_n, X_{\mathcal{I},Q_n}, Y_{\mathcal{I},Q_n}}}(Y_{T^c\cap \mathcal{G}_h,Q_n}|X_{\mathcal{S}^{h},Q_n}, Y_{\mathcal{G}^{h-1},Q_n}, Q_n=k) \notag \\
&\quad = \sum_{h=1}^\alpha H_{\tilde p_{Q_n, X_{\mathcal{I},Q_n}, Y_{\mathcal{I},Q_n}}}(Y_{T^c\cap \mathcal{G}_h,Q_n}|X_{\mathcal{S}^{h},Q_n}, Y_{\mathcal{G}^{h-1},Q_n}, Q_n) \notag \\
&\quad \stackrel{\eqref{MCTimeVariable}}{=} \sum_{h=1}^\alpha H_{\tilde p_{X_{\mathcal{I},Q_n}, Y_{\mathcal{I},Q_n}}}(Y_{T^c\cap \mathcal{G}_h,Q_n}|X_{\mathcal{S}^{h},Q_n},Y_{\mathcal{G}^{h-1},Q_n}). \label{cutseteqnSet2*}
\end{align}
Using \eqref{cutseteqnSet1}, \eqref{cutsetstatement8}, (\ref{cutseteqnSet2}) and
(\ref{cutseteqnSet2*}), we obtain
\begin{align}
  \sum_{i\in T, j\in T^c} \log  M_{i,j} &  \le \sum_{i\in T, j\in T^c} (1 + P_{i,j}^n \log M_{i,j}) \notag\\
   & \quad +  n\sum_{h=1}^\alpha I_{\tilde p_{X_{\mathcal{I},Q_n}, Y_{\mathcal{I},Q_n}}}(X_{T\cap \mathcal{S}^h, Q_n}, Y_{T\cap \mathcal{G}^{h-1},  Q_n} ; Y_{T^c\cap \mathcal{G}_h ,Q_n}| X_{T^c \cap \mathcal{S}^h,Q_n}, Y_{T^c\cap \mathcal{G}^{h-1}, Q_n}), \label{cutseteqnSet5}
\end{align}
where $\tilde p_{X_{\mathcal{I},Q_n}, Y_{\mathcal{I},Q_n}}$ is a distribution on $(\mathcal{X}_{\mathcal{I}}, \mathcal{Y}_{\mathcal{I}})$ that satisfies
\begin{equation*}
\tilde p_{X_{\mathcal{I},Q_n}, Y_{\mathcal{I},Q_n}}(x_{\mathcal{I}}, y_{\mathcal{I}}) \stackrel{\eqref{defTildeP}}{=}\frac{1}{n}\sum_{k=1}^n p_{X_{\mathcal{I},k}, Y_{\mathcal{I},k}}(x_{\mathcal{I}}, y_{\mathcal{I}}).
\end{equation*}
Consider each distribution on $(\mathcal{X}_{\mathcal{I}}, \mathcal{Y}_{\mathcal{I}})$ as a point in an
$|\mathcal{X}_{\mathcal{I}}|| \mathcal{Y}_{\mathcal{I}}|$-dimensional Euclidean space. Let
\[
\{\tilde p_{X_{\mathcal{I},Q_{n_\ell}}, Y_{\mathcal{I},Q_{n_\ell}}}\}_{\ell=1, 2, \ldots}
\]
 be a convergent subsequence
of
\[
\{\tilde p_{X_{\mathcal{I},Q_{n}}, Y_{\mathcal{I},Q_{n}}}\}_{n=1, 2, \ldots}
\]
with respect to the
$\mathcal{L}_1$-distance, where the $\mathcal{L}_1$-distance between
two distributions $u(x)$ and $v(x)$ on the same discrete alphabet
$\mathcal{X}$ is defined as
$\sum_{x\in\mathcal{X}}|u(x)-v(x)|$.
Since the set of all joint distributions on $(\mathcal{X}_\mathcal{I}, \mathcal{Y}_\mathcal{I})$ is closed with respect to the $\mathcal{L}_1$-distance, there
exists a joint distribution $\bar q_{X_{\mathcal{I}}, Y_{\mathcal{I}}}$ such that
\begin{equation}
\bar q_{X_{\mathcal{I}}, Y_{\mathcal{I}}}(x_{\mathcal{I}}, y_{\mathcal{I}}) = \lim_{\ell \rightarrow \infty} \tilde p_{X_{\mathcal{I},Q_{n_\ell}}, Y_{\mathcal{I},Q_{n_\ell}}}(x_{\mathcal{I}}, y_{\mathcal{I}}). \label{marginalQBar}
\end{equation}
For each $h\in \{1, 2, \ldots, \alpha\}$, since
\[
I_{p_{X_{\mathcal{I}}, Y_{\mathcal{I}}}}(X_{T\cap \mathcal{S}^h},Y_{T\cap \mathcal{G}^{h-1}} ;Y_{T^c\cap \mathcal{G}_h}| X_{T^c \cap \mathcal{S}^h}, Y_{T^c\cap \mathcal{G}^{h-1}})
\]
is a continuous functional of $p_{X_{\mathcal{I}}, Y_{\mathcal{I}}}$, it follows from \eqref{marginalQBar} that
\begin{align}
& \lim_{\ell \rightarrow \infty} I_{\tilde p_{X_{\mathcal{I},Q_{n_\ell}}, Y_{\mathcal{I},Q_{n_\ell}}}}(X_{T\cap \mathcal{S}^h, Q_{n_\ell}},Y_{T\cap \mathcal{G}^{h-1}, Q_{n_\ell}} ;Y_{T^c\cap \mathcal{G}_h ,Q_{n_\ell}}| X_{T^c \cap \mathcal{S}^h,Q_{n_\ell}}, Y_{T^c\cap \mathcal{G}^{h-1}, Q_{n_\ell}}) \notag \\
& \: = I_{\bar q_{X_{\mathcal{I}}, Y_{\mathcal{I}}}}(X_{T\cap \mathcal{S}^h},  Y_{T\cap \mathcal{G}^{h-1}} ; Y_{T^c\cap \mathcal{G}_h}| X_{T^c \cap \mathcal{S}^h}, Y_{T^c\cap \mathcal{G}^{h-1}}). \label{eqnContinuousFunctional}
\end{align}
Then,
\begin{align}
 & \!\!\! \sum_{(i,j)\in T\times T^c}R_{i,j} \notag\\
 & \!\! \stackrel{\text{(a)}}{\le}
 \liminf_{n\rightarrow \infty} \sum_{h=1}^\alpha I_{\tilde p_{X_{\mathcal{I},Q_{n}}, Y_{\mathcal{I},Q_{n}}}}(X_{T\cap \mathcal{S}^h, Q_n},Y_{T\cap \mathcal{G}^{h-1}, Q_n} ;Y_{T^c\cap \mathcal{G}_h ,Q_n}|  X_{T^c \cap \mathcal{S}^h,Q_n}, Y_{T^c\cap \mathcal{G}^{h-1}, Q_n}) \notag\\
  &\!\! \le
  \lim_{\ell \rightarrow \infty}\sum_{h=1}^\alpha I_{\tilde p_{X_{\mathcal{I},Q_{n_\ell}}, Y_{\mathcal{I},Q_{n_\ell}}}}(X_{T\cap \mathcal{S}^h, Q_{n_\ell}},Y_{T\cap \mathcal{G}^{h-1}, Q_{n_\ell}} ;Y_{T^c\cap \mathcal{G}_h ,Q_{n_\ell}}| X_{T^c \cap \mathcal{S}^h,Q_{n_\ell}}, Y_{T^c\cap \mathcal{G}^{h-1}, Q_{n_\ell}}) \notag\\
  &\!\! \stackrel{\eqref{eqnContinuousFunctional}}{=}  \! \sum_{h=1}^\alpha  I_{\bar q_{X_{\mathcal{I}}, Y_{\mathcal{I}}}}(X_{T\cap \mathcal{S}^h}, \! Y_{T\cap \mathcal{G}^{h-1}} ; \! Y_{T^c\cap \mathcal{G}_h}| X_{T^c \cap \mathcal{S}^h},\! Y_{T^c\cap \mathcal{G}^{h-1}}\!),
\label{cutsetR1one*}
\end{align}
where (a) follows from \eqref{cutseteqnSet5}, \eqref{thmTempEq1} and \eqref{thmTempEq2}.
Define
\[
q^{(h, n)}_{X_{\mathcal{S}^h}, Y_{\mathcal{G}^{h-1}}}(x_{\mathcal{S}^h}, y_{\mathcal{G}^{h-1}})=\frac{1}{n}\sum_{k=1}^n p_{X_{\mathcal{S}^h,k}, Y_{\mathcal{G}^{h-1},k} } (x_{\mathcal{S}^h}, y_{\mathcal{G}^{h-1}})
\]
for each $h\in \{1, 2, \ldots, \alpha\}$. Then, $ \bar q_{X_{\mathcal{S}^h}, Y_{\mathcal{G}^h}}$
(the marginal distribution of $\bar q_{X_{\mathcal{I}}, y_{\mathcal{I}}}$) satisfies
\begin{align}
& \bar q_{X_{\mathcal{S}^h}, Y_{\mathcal{G}^h}}(x_{\mathcal{S}^h}, y_{\mathcal{G}^h}) \notag \\
  &\quad \stackrel{\eqref{marginalQBar}}{=} \lim_{\ell \rightarrow \infty} p_{X_{\mathcal{S}^h,Q_{n_\ell}}, Y_{\mathcal{G}^h,Q_{n_\ell}}}(x_{\mathcal{S}^h},
y_{\mathcal{G}^h})\notag \\
 & \quad \stackrel{\text{(a)}}{=}  \lim_{\ell \rightarrow \infty}  q^{(h,n_\ell)}_{X_{\mathcal{S}^h}, Y_{\mathcal{G}^{h-1}}}(x_{\mathcal{S}^h}, y_{\mathcal{G}^{h-1}}) p_{Y_{\mathcal{G}_h} | X_{\mathcal{S}^h},Y_{\mathcal{G}^{h-1}}}^{(h)}(y_{\mathcal{G}_h} | x_{\mathcal{S}^h},y_{\mathcal{G}^{h-1}}) \label{MIstatement1}
 \end{align}
where (a) follows from summing over all $k$ on both sides of \eqref{cutsetstatement12}.
In addition, it follows from summing over all $y_{\mathcal{G}_h}$ on both sides of \eqref{MIstatement1} that $\bar q_{X_{\mathcal{S}^h}, Y_{\mathcal{G}^{h-1}}}$ satisfies
\begin{equation}
\bar q_{X_{\mathcal{S}^h}, Y_{\mathcal{G}^{h-1}}}(x_{\mathcal{S}^h}, y_{\mathcal{G}^{h-1}}) = \lim_{\ell \rightarrow \infty} q^{(h,n_\ell)}_{X_{\mathcal{S}^h}, Y_{\mathcal{G}^{h-1}}}(x_{\mathcal{S}^h}, y_{\mathcal{G}^{h-1}}). \label{MIstatement2}
\end{equation}
Then,
 \begin{align}
\bar q_{X_{\mathcal{S}^h}, Y_{\mathcal{G}^h}}
 & \,\stackrel{\text{(a)}}{=}  \bar q_{X_{\mathcal{S}^h}, Y_{\mathcal{G}^{h-1}}}q_{Y_{\mathcal{G}_h} | X_{\mathcal{S}^h},Y_{\mathcal{G}^{h-1}}}^{(h)}\notag \\
 &\, = \bar q_{X_{\mathcal{S}^{h-1}}, Y_{\mathcal{G}^{h-1}}}\bar q_{X_{\mathcal{S}_h}|X_{\mathcal{S}^{h-1}}, Y_{\mathcal{G}^{h-1}}}   q_{Y_{\mathcal{G}_h}  |X_{\mathcal{S}^h}  ,  Y_{\mathcal{G}^{h-1}}}^{(h)} \label{eqnMarginalDistributionRecursion}
\end{align}
for each $h\in \{1, 2, \ldots, \alpha\}$, where (a) follows from \eqref{MIstatement1} and \eqref{MIstatement2}. Consequently,
\begin{align}
 \bar q_{X_{\mathcal{I}}, Y_{\mathcal{I}}}
&  \stackrel{\text{(a)}}{=}  \bar q_{X_{\mathcal{S}^\alpha}, Y_{\mathcal{G}^\alpha}} \notag \\
& \stackrel{\text{(b)}}{=}  \prod_{h=1}^\alpha  \bar q_{X_{\mathcal{S}_h}|X_{\mathcal{S}^{h-1}}, Y_{\mathcal{G}^{h-1}} }q_{Y_{\mathcal{G}_h}  |X_{\mathcal{S}^h}  ,  Y_{\mathcal{G}^{h-1}}}^{(h)}  \label{theoremStatement2}
\end{align}
where
\begin{enumerate}
\item[(a)] follows from the fact that $\cup_{h=1}^\alpha \mathcal{S}_h = \cup_{h=1}^\alpha \mathcal{G}_h = \mathcal{I}$.
\item[(b)] follows from \eqref{eqnMarginalDistributionRecursion} by recursion.
\end{enumerate}
Since $\bar q_{X_{\mathcal{I}}, Y_{\mathcal{I}}}$ depends only on the sequence of $(B, n, M_{\mathcal{I}\times \mathcal{I}})$-codes but not on $T$, the theorem follows from (\ref{cutsetR1one*}) and (\ref{theoremStatement2}).
\end{IEEEproof}

\section{Capacity Region of BSC with Correlated Feedback}\label{sectionGeneralized2way}
Let $\mathcal{C}$ denote the capacity region of the BSC with correlated feedback in Example~\ref{exampleTwoNode} (cf.\ Figure~\ref{BSCFB}).
Let $H(\epsilon)$ denote the entropy of a Bernoulli random variable $X$ with $\Pr\{X=0\} = \epsilon$. It then follows from Theorem~\ref{cutsetThmCapacity} that for each achievable rate tuple $R_{\{1, 2\}\times\{1, 2\}}$, there exists some $p_{X_1, X_2, Y_1, Y_2}$ such that
\begin{align}
R_{1,2} & \le I_{p_{X_1, Y_2}}(X_1;Y_2) \notag \\
& \le 1-H_{p_{X_1, Y_2}}(Y_2|X_1) \notag \\
& \stackrel{\eqref{definitionP1}}{=} 1-H(\epsilon) \label{firstInequalityExample}
\end{align}
and
\begin{align}
R_{2,1}  &\le I_{p_{X_1, X_2, Y_1, Y_2}}(X_2, Y_2;Y_1|X_1) \notag \\
 & \le 1- H_{p_{X_1, X_2, Y_1, Y_2}}(Y_1|X_1, X_2, Y_2) \notag \\
 & \stackrel{\eqref{definitionP2}}{=} 1. \label{secondInequalityExample}
\end{align}
Let
\begin{equation*}
\mathcal{R}^*\! = \!\left\{\parbox[c]{1.35 in}{$(0, R_{1,2}, R_{2,1}, 0) \in \mathbb{R}_+^4$}\!\!\left| \!\! \begin{array}{c}R_{1,2}  \le 1-H(\epsilon) ,\\ R_{2,1} \le 1
\end{array}\right. \!\!\! \!\right\}\!.
\end{equation*}
It then follows from \eqref{firstInequalityExample} and \eqref{secondInequalityExample} that
\begin{equation}
\mathcal{C} \subseteq \mathcal{R}^*. \label{cutset10**}
\end{equation}

 Fix any $\delta>0$. Consider a capacity-achieving block code of length $n$ for the BSC with crossover probability $\epsilon$ with rate
\begin{equation}
R_{1,2}\le 1-H(\epsilon)- \delta. \label{rateArbitrarilyClose}
 \end{equation}
 Such a code encodes the message $W_{1,2}$ that is uniformly distributed on
 \begin{equation}
 \{1, 2, \ldots, \lceil 2^{n R_{1,2}} \rceil\} \label{rateArbitrarilyClose2}
  \end{equation}
  into a codeword consisting of a sequence of bits $\{X_{1,k}^\prime\}_{k=1, 2, \ldots, n}$. In the $k^{\text{th}}$ time slot, node~1 transmits
  \begin{equation}
  X_{1,k}=X_{1,k}^\prime \label{rateArbitrarilyClose3}
  \end{equation}
  through channel $q^{(1)}$.
The message $W_{2,1}$ consists of a sequence of $n$ uniform i.i.d.\ bits $\{X_{2,k}^\prime\}_{k=1, 2, \ldots, n}$. In the $k^{\text{th}}$ time slot, upon receiving $Y_{2,k}$, node~2 transmits
\begin{equation}
X_{2,k}=X_{2,k}^\prime +Y_{2,k} \label{rateArbitrarilyClose4}
\end{equation}
 through channel $q^{(2)}$, whose output bit $Y_{1,k}$ is received by node~1.

 Since
\[
\Pr\{Y_{2,k}=X_{1,k}^\prime\}=1-\epsilon
 \]
by \eqref{definitionP1} and the capacity of the BSC with crossover probability~$\epsilon$ is $1-H(\epsilon)$, it follows from \eqref{rateArbitrarilyClose}, \eqref{rateArbitrarilyClose2} and \eqref{rateArbitrarilyClose3} that node~2 can decode $W_{1,2}$ with vanishing probability of error as $n$ goes to infinity. Since $\delta$ is arbitrary, node~1 can transmit $W_{1,2}$ at a rate arbitrarily close to $1-H(\epsilon)$ such that node~2 can decode $W_{1,2}$ with probability approaching~1 as $n \rightarrow \infty$.
On the other hand, since
\[
\Pr\{Y_{1,k} = X_{2,k} + Y_{2, k}\}=1
\]
  by \eqref{definitionP2}, it follows that with probability~1,
\begin{align*}
Y_{1,k} & = X_{2,k} + Y_{2, k} \\
& \stackrel{\eqref{rateArbitrarilyClose4}}{=} (X_{2,k}^\prime +Y_{2,k})  + Y_{2, k}\\
&  =X_{2,k}^\prime.
\end{align*}
Therefore, node~1 receives the bit sequence $\{X_{2,k}^\prime\}_{k=1, 2, \ldots, n}$ without error with probability one for any~$n$.
Consequently, $(0,1-H(\epsilon), 1,0)$ is achievable, which implies from \eqref{cutset10**} that $\mathcal{C} = \mathcal{R}^*$.

\section{Proof of Cut-Set Bound on Positive-Delay Region}\label{sectionOuterBound*}
In this section, we will provide a proof of Theorem~\ref{cutsetThmPositive}. To this end, it suffices to prove the following folklore theorem, whose proof is tedious and therefore relegated to the appendix.
\medskip
\begin{Theorem} \label{thmEquivalentNetwork}
For any $(\boldsymbol{1}, n, M_{\mathcal{I}\times\mathcal{I}})$-code, a DMN specified by $(\mathcal{X}_\mathcal{I}, \mathcal{Y}_\mathcal{I}, \alpha, \boldsymbol{\mathcal{S}}, \boldsymbol{\mathcal{G}}, \boldsymbol{q})$ is equivalent to a DMN specified by $(\mathcal{X}_\mathcal{I}, \mathcal{Y}_\mathcal{I}, 1, \mathcal{I}, \mathcal{I}, q^{(1)}q^{(2)} \ldots q^{(\alpha)})$.
\end{Theorem}
\medskip
To facilitate discussion, we rewrite Theorem~\ref{cutsetThmCapacity} in a slightly different way to prove the cut-set bound on $\mathcal{R}_{\boldsymbol{1}}$ in Theorem~\ref{cutsetThmPositive}.
\medskip
\begin{Theorem}[Identical to Theorem~\ref{cutsetThmCapacity}]\label{theoremCited}
Let $(\mathcal{X}_\mathcal{I}, \mathcal{Y}_\mathcal{I}, \gamma, \boldsymbol{\mathcal{S}}, \boldsymbol{\mathcal{G}}, (\tilde q^{(1)}, \tilde  q^{(2)}, \ldots, \tilde  q^{(\gamma)}))$ be a DMN. Then for each achievable rate tuple $R_{\mathcal{I}\times \mathcal{I}}$, there exists a joint distribution $p_{X_\mathcal{I}, Y_\mathcal{I}}$ satisfying
\[
  p_{X_\mathcal{I}, Y_\mathcal{I}} =
 \prod_{h=1}^\gamma p_{X_{\mathcal{S}_h}|X_{\mathcal{S}^{h-1}}, Y_{\mathcal{G}^{h-1}}} \tilde q_{Y_{\mathcal{G}_h} | X_{\mathcal{S}^h},  Y_{\mathcal{G}^{h-1}}}^{(h)}
\]
such that for any $T\subseteq \mathcal{I}$,
\begin{align*}
 \sum_{ (i,j)\in T\times T^c} R_{i,j}  \le  \sum_{h=1}^\gamma  I_{p_{X_\mathcal{I}, Y_\mathcal{I}}}(X_{T\cap \mathcal{S}^h}, Y_{T\cap \mathcal{G}^{h-1}} ;  Y_{T^c\cap \mathcal{G}_h}| X_{T^c \cap \mathcal{S}^h},  Y_{T^c\cap \mathcal{G}^{h-1}}).
\end{align*}
\end{Theorem}
\medskip
The proof of Theorem~\ref{cutsetThmPositive} is a direct consequence of Theorems~\ref{thmEquivalentNetwork} and~\ref{theoremCited}, which is shown as follows for completeness.
\begin{IEEEproof}[\textbf{Proof of Theorem~\ref{cutsetThmPositive}}]
Since $(\mathcal{X}_\mathcal{I}, \mathcal{Y}_\mathcal{I}, \alpha, \boldsymbol{\mathcal{S}}, \boldsymbol{\mathcal{G}}, \boldsymbol{q})$ is equivalent to $(\mathcal{X}_\mathcal{I}, \mathcal{Y}_\mathcal{I}, 1, \mathcal{I}, \mathcal{I}, q^{(1)} q^{(2)} \ldots q^{(\alpha)})$ for each $(\boldsymbol{1}, n, M_{\mathcal{I}\times\mathcal{I}})$-code by Theorem~\ref{thmEquivalentNetwork}, it follows from Theorem~\ref{theoremCited} by setting $\gamma=1$, $\mathcal{S}^1=\mathcal{G}^1=\mathcal{I}$ and $\tilde q^{(1)} = q^{(1)} q^{(2)}\ldots q^{(\alpha)}$ that for each $\boldsymbol{1}$-achievable rate tuple $R_{\mathcal{I}\times \mathcal{I}}$, there exists a joint distribution $p_{X_\mathcal{I}, Y_\mathcal{I}}$ satisfying
\[
  p_{X_\mathcal{I}, Y_\mathcal{I}} =
  p_{X_{\mathcal{I}}} \prod_{h=1}^\alpha q_{Y_{\mathcal{G}_h} | X_{\mathcal{S}^h},  Y_{\mathcal{G}^{h-1}}}^{(h)}
\]
such that for any $T\subseteq \mathcal{I}$,
\begin{align*}
 \sum_{ (i,j)\in T\times T^c} R_{i,j}  \le  I_{ p_{X_\mathcal{I}, Y_\mathcal{I}}}(X_T, Y_T; X_{T^c}).
\end{align*}
Consequently, $\mathcal{R}_\mathbf{1} \subseteq \mathcal{R}_\mathbf{1}^{\text{out}}$ (cf.\ \eqref{RoutPositive}).
\end{IEEEproof}

\section{Positive-Delay Region Strictly Smaller than Capacity Region}\label{illustratingExample}
For some generalized DMN, the positive-delay region can be strictly smaller than the capacity region. This has been demonstrated by El~Gamal \textit{et al}.\ \cite[Section IV]{AbbasRelayNetwork} for the relay-without-delay channel which consists of three nodes. In this section, we demonstrate the same for a two-node network.

Let $\mathcal{C}$ denote the capacity region of the BSC with correlated feedback (cf.\ Figure~\ref{BSCFB}). It is shown in Section~\ref{sectionGeneralized2way} that
 \begin{equation}
\mathcal{C}
 =  \left\{  \parbox[c]{1.35 in}{$(0, R_{1,2}, R_{2,1}, 0) \in \mathbb{R}_+^4$}  \left| \begin{array}{c}R_{1,2}  \le 1-H(\epsilon) ,\\ R_{2,1} \le 1
\end{array}\right. \!\!\right\}. \label{exampleCapacity}
\end{equation}
In the rest of this section, we will show that $\mathcal{R}_{(1, 1)} \subsetneq \mathcal{C}$.

Let $\mathcal{R}_{(1, 1)}^{\text{out}}$ denote
\begin{equation}
\bigcup_{\substack{p_{X_{\{1,2\}}, Y_{\{1, 2\}}}:p_{X_{\{1,2\}}, Y_{\{1, 2\}}}=\\ p_{X_1, X_2} q_{Y_2|X_1}^{(1)} q_{Y_1|X_1, X_2, Y_2}^{(2)}}}
\left\{\parbox[c]{1.38 in}{$(0, R_{1,2},  R_{2,1}, 0)\in \mathbb{R}_+^4$} \left| \! \begin{array}{c}R_{1,2}  \le I_{p_{X_{\{1,2\}}, Y_{\{1, 2\}}}}(X_{1};Y_{2}|X_2),\\ R_{2,1} \le
 I_{p_{X_{\{1,2\}}, Y_{\{1, 2\}}}}(X_{2};Y_{1}|X_{1})
\end{array}\right. \right\}.
\label{cutset11}
\end{equation}
It then follows from Theorem~\ref{cutsetThmPositive} that
 \begin{equation}
 \mathcal{R}_{(1, 1)} \subseteq \mathcal{R}_{(1, 1)}^{\text{out}}. \label{exampleEquation5}
\end{equation}
\medskip
For any $p_{X_{\{1,2\}}, Y_{\{1, 2\}}}$ distributed according to
\begin{equation*}
p_{X_{\{1,2\}}, Y_{\{1, 2\}}}= p_{X_1, X_2} q_{Y_2|X_1}^{(1)} q_{Y_1|X_1, X_2, Y_2}^{(2)}, 
\end{equation*}
since the marginal distribution $p_{X_{\{1,2\}}, Y_2}$ satisfies
\[
p_{X_{\{1,2\}}, Y_2}= p_{X_1, X_2} q_{Y_1|X_1, X_2, Y_2}^{(2)},
\]
it follows from Proposition~\ref{propositionCutsetMCLemma*} that
\[
(X_2 \rightarrow X_1 \rightarrow Y_2)_p 
\]
forms a Markov Chain, which implies that
\begin{align}
H_{p_{X_{\{1,2\}}, Y_{\{1, 2\}}}}(Y_2|X_1, X_2) &= H_{p_{X_{\{1,2\}}, Y_{\{1, 2\}}}}(Y_2|X_1) \notag \\
& = H(\epsilon) \label{exampleTempTempEquation}
\end{align}
where the last equality follows from \eqref{definitionP1}. In addition, omitting subscripts $p_{X_{\{1,2\}}, Y_{\{1, 2\}}}$ for the entropy and mutual information terms, we have
\begin{align}
I(X_1;Y_2|X_2)
& = H(Y_2|X_2)- H(Y_2|X_1, X_2)\notag \\
& \le 1 -  H(Y_2|X_1, X_2)\label{exampleLastEqu1}
\end{align}
and
 \begin{align}
 I(X_{2};Y_{1}|X_{1})
 & = H(Y_1|X_1) - H(Y_1|X_1, X_2) \notag\\
 & \le 1 - H(Y_1|X_1, X_2) \notag\\
 & \stackrel{\eqref{definitionP2}}{=} 1 - H(X_2+Y_2|X_1, X_2)\notag\\
 & = 1 - H(Y_2 | X_1, X_2).  \label{exampleLastEqu2}
 \end{align}
 It then follows from \eqref{cutset11}, \eqref{exampleLastEqu1}, \eqref{exampleLastEqu2} and \eqref{exampleTempTempEquation} that
\begin{equation*}
 \mathcal{R}_{(1, 1)}^{\text{out}} \subseteq \left\{\parbox[c]{1.36 in}{$(0, R_{1,2}, R_{2,1}, 0) \in \mathbb{R}_+^4$}\left| \begin{array}{c}R_{1,2}  \le 1-H(\epsilon) ,\\ R_{2,1} \le 1-H(\epsilon)
\end{array}\right. \!\!\! \!\right\}\!,
\end{equation*}
 which implies from \eqref{exampleCapacity} that
$
 \mathcal{R}_{(1, 1)}^{\text{out}} \subsetneq \mathcal{C}
$
 for any $0<\epsilon<1$, which then implies from
\eqref{exampleEquation5} that
\begin{equation}
\mathcal{R}_{(1, 1)} \subsetneq \mathcal{C} \label{eqnR11<C}
\end{equation}
for any $0<\epsilon<1$.
\medskip
\begin{Remark}
The intuition behind \eqref{eqnR11<C} is as follows: Since the noises in the forward and reverse links are correlated as shown in Fig.~\ref{BSCFB} and node~2 incurs no delay, node~2 can employ some sort of ``dirty-paper coding" as described in Section~\ref{sectionGeneralized2way} so that the noise in the signal received by node~1 can be neutralized. In contrast, if node~2 incurs a delay, then it cannot use the signal received in a time slot to neutralize the noise incurred on the reverse link in the same time slot, resulting in a lower capacity in the reverse link compared with the case when node~2 incurs no delay.
\end{Remark}
\section{Causal Relay Network} \label{sectionRelayWithoutDelay}
The causal relay network \cite{causalRelayNetwork} is a generalization of the relay-without-delay channel \cite{AbbasRelayNetwork}. In this section, we demonstrate that the causal relay network is a special case of the generalized DMN. The causal relay network consists of a set of nodes that incur no delay, denoted by $\mathcal{N}_0$, and a set of nodes that incur a delay, denoted by $\mathcal{N}_1$. The causal relay network is specified by the following two channels: $q_{Y_{\mathcal{N}_0}|X_{\mathcal{N}_1}}^{(1)}$ and $q_{Y_{\mathcal{N}_1}|X_{\mathcal{N}_1}, X_{\mathcal{N}_0}, Y_{\mathcal{N}_0}}^{(2)}$. For any $(B, n, M_{\mathcal{I}\times \mathcal{I}})$-code (cf.\ Definition~\ref{defCode}) on the causal relay network, $X_{\mathcal{I},k}$ and $Y_{\mathcal{I},k}$ are generated in the $k^{\text{th}}$ time slot in the order
\[
X_{\mathcal{N}_1,k}, Y_{\mathcal{N}_0,k}, X_{\mathcal{N}_0,k}, Y_{\mathcal{N}_1,k}.
 \]
 Therefore, the discrete memoryless causal relay network formulated in \cite{causalRelayNetwork} is the same as the generalized DMN $(\mathcal{X}_{\mathcal{I}}, \mathcal{Y}_{\mathcal{I}}, 2, (\mathcal{N}_1, \mathcal{N}_0), (\mathcal{N}_0, \mathcal{N}_1), (q^{(1)}, q^{(2)}))$ (cf.\ Definition~\ref{defDiscreteNetwork}). For each achievable rate tuple $R_{\mathcal{I}\times \mathcal{I}}$ for this network, it follows from Theorem~\ref{cutsetThmCapacity} that there exists a joint distribution $p_{X_{\mathcal{I}}, Y_{\mathcal{I}}}$ satisfying
\begin{align*}
 p_{X_{\mathcal{I}}, Y_{\mathcal{I}}} =  p_{X_{\mathcal{N}_1}} q_{Y_{\mathcal{N}_0}|X_{\mathcal{N}_1}}^{(1)}
 p_{X_{\mathcal{N}_0}|X_{\mathcal{N}_1},Y_{\mathcal{N}_0}} q_{Y_{\mathcal{N}_1}|X_{\mathcal{N}_1}, X_{\mathcal{N}_0}, Y_{\mathcal{N}_0}}^{(2)}
\end{align*}
 such that for any $T\subseteq \mathcal{I}$,
\begin{align*}
 \sum_{ i\in T, j\in T^c} R_{i,j}
 \le  I_{p_{X_{\mathcal{I}}, Y_{\mathcal{I}}}}(X_{T\cap \mathcal{N}_1};  Y_{T^c \cap \mathcal{N}_0}| X_{T^c \cap \mathcal{N}_1}) + I_{p_{X_{\mathcal{I}}, Y_{\mathcal{I}}}}(X_{T}, Y_{T\cap \mathcal{N}_0} ;  Y_{T^c\cap \mathcal{N}_1}| X_{T^c},  Y_{T^c\cap \mathcal{N}_0}),
\end{align*}
which recovers the cut-set bound for the causal relay network stated in Theorem~1 of \cite{causalRelayNetwork}.

In addition, if $\mathcal{N}_0=\emptyset$, then every node incurs a delay, which implies from Theorem~\ref{cutsetThmPositive} that for each $\boldsymbol{1}$-achievable rate tuple $R_{\mathcal{I}\times \mathcal{I}}^*$, there exists a joint distribution $p_{X_{\mathcal{I}}, Y_{\mathcal{I}}}$ satisfying
\begin{equation*}
 p_{X_{\mathcal{I}}, Y_{\mathcal{I}}}= p_{X_{\mathcal{I}}} q_{Y_{\mathcal{N}_0}|X_{\mathcal{N}_1}}^{(1)}
 q_{Y_{\mathcal{N}_1}|X_{\mathcal{N}_1}, X_{\mathcal{N}_0}, Y_{\mathcal{N}_0}}^{(2)}
\end{equation*}
such that for any $T\subseteq \mathcal{I}$,
\begin{equation*}
\sum_{i\in T, j\in T^c}R_{i,j}^* \le I_{p_{X_{\mathcal{I}}, Y_{\mathcal{I}}}}(X_{T};Y_{T^c}| X_{T^c}).
\end{equation*}
We end this section by demonstrating that for some Gaussian causal relay network, the positive-delay region is strictly smaller than the capacity region.
\medskip
\begin{Example} \label{exampleCausalRelayNetwork}
Consider a Gaussian causal relay network consisting of three nodes as illustrated in Figure~\ref{GaussianCausalRelay}, where node~1 wants to transmit a message to node~3 with the help of node~2 in~$n$ time slots. \begin{figure}[!t]
\center
\includegraphics[width=2.5 in, height=1.3 in, bb = 203 287 631 495, angle=0, clip=true]{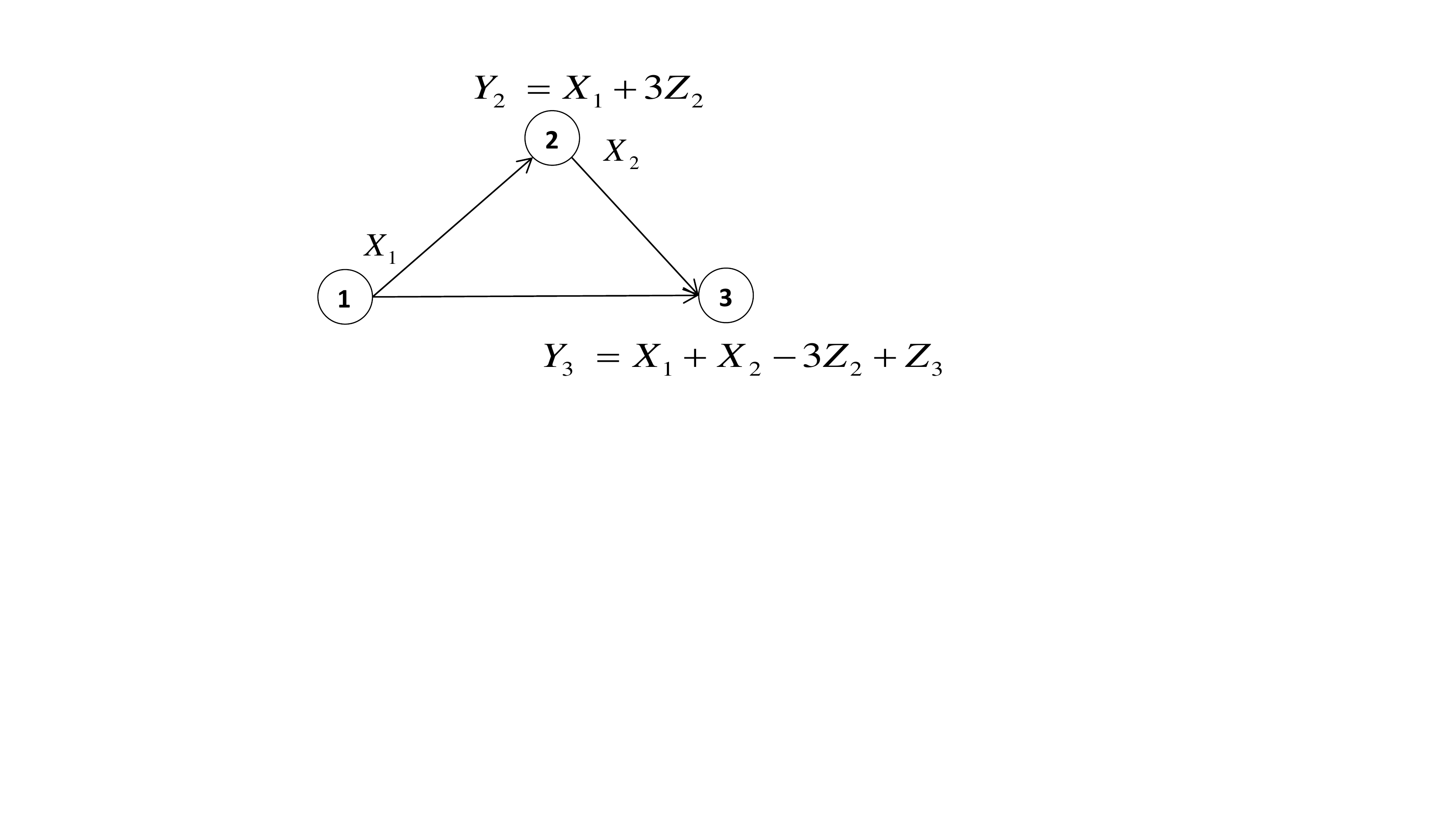}
\caption{A Gaussian causal relay network.}
\label{GaussianCausalRelay}
\end{figure} In every time slot~$k$, node~1 and node~2 transmit $X_{1,k} \in \mathbb{R}$ and $X_{2,k} \in \mathbb{R}$ respectively, and node~2 and node~3 receive $Y_{2,k}\in \mathbb{R}$ and $Y_{3,k}\in \mathbb{R}$ respectively. The network is characterized by channels $q_{Y_2|X_1}^{(1)}$ and $q_{Y_3|X_1, X_2, Y_2}^{(2)}$ which are readily determined by the following two statements for each $k\in\{1, 2, \ldots, n\}$:
\begin{equation}
\Pr\{Y_{2,k} = x_{1,k} + 3 Z_{2,k}\}=1 \label{probability1Example2}
\end{equation}
for all $x_{1,k}$, and
\begin{equation}
\Pr\{Y_{3,k} = 2x_{1,k} + x_{2,k} - y_{2,k} + Z_{3,k} \}=1 \label{probability2Example2}
\end{equation}
for all $x_{1,k}$, $x_{2,k}$ and $y_{2,k}$,
where $Z_{2,k}$ and $Z_{3,k}$ are independent standard normal random variables and $\{(Z_{2,k}, Z_{3,k})\}_{k=1}^n$ are independent. In addition, let~$P$ and $P+10$ be the admissible power for node~1 and node~2 respectively such that
\begin{equation}
\Pr\left\{\sum_{k=1}^n X_{1,k}^2 \le n P\right\}=1 \label{powerConstraintExample2}
\end{equation}
and
\begin{equation}
\Pr\left\{\sum_{k=1}^n X_{2,k}^2 \le n (P+10)\right\}=1. \label{powerConstraintExample2*}
\end{equation}
The Gaussian version of Theorem~\ref{cutsetThmPositive} implies that for every $\mathbf{1}$-achievable rate~$R_\mathbf{1}$,
there exists a probability distribution $p_{X_{1,k}, X_{2,k}, Y_{2,k}, Y_{3,k}}$ satisfying
\[
p_{X_{1,k}, X_{2,k}, Y_{2,k}, Y_{3,k}}(x_{1,k}, x_{2,k}, y_{2,k}, y_{3,k}) = p_{X_{1,k}, X_{2,k}}(x_{1,k}, x_{2,k}) q_{Y_{2}|X_{1}}^{(1)}(y_{2,k}|x_{1,k}) q_{Y_{3}|X_{1}, X_{2}, Y_{2}}^{(2)}(y_{3,k}|x_{1,k}, x_{2,k}, y_{2,k})
\]
for each $k\in\{1, 2, \ldots, n\}$ such that
\begin{equation}
R_\mathbf{1}\le \frac{1}{n}\sum_{k=1}^n I_{p_{X_{1,k}, X_{2,k}, Y_{2,k}, Y_{3,k}}}(X_{1,k}, X_{2,k};Y_{3,k}). \label{eqn1Example2}
\end{equation}
Under the probability distribution $p_{X_{1,k}, X_{2,k}, Y_{2,k}, Y_{3,k}}$, it follows from \eqref{probability1Example2} and \eqref{probability2Example2} that
\begin{equation}
\Pr\{Y_{2,k} = X_{1,k} + 3Z_{2,k}\}=1 \label{probability3Example2}
\end{equation}
and
\begin{equation}
\Pr\{Y_{3,k} = X_{1,k} + X_{2,k} -3 Z_{2,k} + Z_{3,k} \}=1 \label{probability4Example2}
\end{equation}
respectively (cf.\ Figure~\ref{GaussianCausalRelay}). Following \eqref{eqn1Example2} and omitting the subscripts for the entropy and mutual information terms, we consider
\begin{align}
& \frac{1}{n}\sum_{k=1}^n I(X_{1,k}, X_{2,k};Y_{3,k}) \notag\\
& = \frac{1}{n}\sum_{k=1}^n \left( h(Y_{3,k}) - h(Y_{3,k}|X_{1,k}, X_{2,k})\right)\notag\\
& \stackrel{\eqref{probability4Example2}}{=} \frac{1}{n}\sum_{k=1}^n \left( h(Y_{3,k}) - h(- 3Z_{2,k} + Z_{3,k}|X_{1,k}, X_{2,k})\right)\notag\\
& \stackrel{\text{(a)}}{=} \frac{1}{n}\sum_{k=1}^n \left( h(Y_{3,k}) - h(- 3Z_{2,k} + Z_{3,k})\right)\notag\\
& \stackrel{\text{(b)}}{=} \frac{1}{n}\sum_{k=1}^n \left( h(Y_{3,k}) - \log_2 \sqrt{20\pi e}\right)\notag\\
& \stackrel{\text{(c)}}{\le} \frac{1}{n}\sum_{k=1}^n \left( \log_2 \sqrt{2\pi e \Var\left[Y_{3,k}\right]} - \log_2 \sqrt{20\pi e}\right)\notag\\
& \stackrel{\text{(d)}}{\le} \frac{1}{n}\sum_{k=1}^n \left( \log_2 \sqrt{2\pi e \left(\Var\left[X_{1,k}+ X_{2,k}\right]+10\right)} - \log_2 \sqrt{20\pi e}\right)\notag\\
& = \frac{1}{n}\sum_{k=1}^n \frac{1}{2} \log_2 \left(1+\frac{\Var\left[X_{1,k}+ X_{2,k}\right]}{10}\right)\notag\\
& \le \frac{1}{n}\sum_{k=1}^n \frac{1}{2} \log_2 \left(1+\frac{\E\left[(X_{1,k}+ X_{2,k})^2\right]}{10}\right)\notag\\
& \le \frac{1}{n}\sum_{k=1}^n \frac{1}{2} \log_2 \left(1+\frac{\E\left[2(X_{1,k}^2+ X_{2,k}^2)\right]}{10}\right)\notag\\
& \stackrel{\text{(e)}}{\le} \frac{1}{2} \log_2 \left(1+\frac{\frac{1}{n}\sum_{k=1}^n\E\left[2(X_{1,k}^2+ X_{2,k}^2)\right]}{10}\right) \notag\\
& \stackrel{\text{(f)}}{\le}  \frac{1}{2} \log_2 \left(3+\frac{2P}{5}\right) \label{eqn2Example2}
\end{align}
where
\begin{enumerate}
\item[(a)] follows from the fact that $(X_{1,k}, X_{2,k})$ and $(Z_{2,k},Z_{3,k})$ are independent.
\item[(b)] follows from the facts that $- 3Z_{2,k} + Z_{3,k}$ is a Gaussian random variable whose mean and variance are $0$ and $10$ respectively and whose differential entropy is $\log_2 \sqrt{20\pi e}$.
    \item[(c)] follows from the fact that the differential entropy of a random variable with variance~$\sigma^2$ is upper bounded by the differential entropy of a Gaussian random variable with variance~$\sigma^2$, which is equal to $\log_2 \sqrt{2\pi \sigma^2}$.
        \item[(d)] follows from \eqref{probability4Example2} and the fact that $(X_{1,k}, X_{2,k})$, $Z_{2,k}$ and $Z_{3,k}$ are independent.
        \item[(e)] follows from Jensen's inequality.
        \item[(f)] follows from \eqref{powerConstraintExample2} and \eqref{powerConstraintExample2*}.
\end{enumerate}
Combining \eqref{eqn1Example2} and \eqref{eqn2Example2}, we have
\begin{equation}
R_\mathbf{1}\le \frac{1}{2} \log_2 \left(3+\frac{2P}{5}\right) \label{eqn6Example2}
\end{equation}
for every $\mathbf{1}$-achievable rate $R_\mathbf{1}$.

In the rest of this example, we are going to show that a higher rate than \eqref{eqn6Example2} can be achieved if node~2 incurs no delay. Suppose node~2 incurs no delay. Then node~2 can send
\begin{align}
X_{2,k} & = \begin{cases}Y_{2,k} &\text{if $\sum_{\ell=1}^k Y_{2,\ell}^2\le n(P+10)$,}\\0 & \text{otherwise} \end{cases} \notag\\
&\stackrel{\eqref{probability1Example2}}{=} \begin{cases}X_{1,k} + 3Z_{2,k} &\text{if $\sum_{\ell=1}^k Y_{2,\ell}^2\le n(P+10)$,} \\0  & \text{otherwise}\end{cases}  \label{eqn3Example2}
\end{align}
so that the power constraint \eqref{powerConstraintExample2*} is always satisfied and
\begin{align}
Y_{3,k} & \stackrel{\eqref{probability4Example2}}{=}\begin{cases} X_{1,k} + X_{2,k} - 3Z_{2,k} + Z_{3,k}&\text{if $\sum_{\ell=1}^k Y_{2,\ell}^2\le n(P+10)$,}\\ X_{1,k}- 3Z_{2,k} + Z_{3,k}& \text{otherwise}\end{cases} \notag\\
& \stackrel{\eqref{eqn3Example2}}{=} \begin{cases} 2X_{1,k} + Z_{3,k} &\text{if $\sum_{\ell=1}^k Y_{2,\ell}^2\le n(P+10)$,}\\ X_{1,k}- 3Z_{2,k} + Z_{3,k}& \text{otherwise.}\end{cases} \label{eqn4Example2}
\end{align}
For any $\delta>0$, there exists by standard channel coding arguments a sequence of Gaussian codebooks with power $P-\delta$ for the Gaussian channel specified by $Y_{3}^n = 2X_{1}^n + Z_{3}^n$ such that the rate of the codebook achieves $\frac{1}{2}\log(1+2(P-\delta))$ with vanishing error probability as the blocklength $n$ increases,
 \begin{equation}
\lim_{n\rightarrow \infty}\Pr\left\{\sum_{k=1}^n X_{1,k}^2\le nP \right\}=1 \label{eqn5Example2}
 \end{equation}
 and
 \begin{equation}
\lim_{n\rightarrow \infty}\Pr\left\{\sum_{k=1}^n (X_{1,k}+3 Z_{2,k})^2\le n(P+10)\right\}=1 \label{eqn5Example2}
 \end{equation}
 (recall that $\{Z_{2,k}\}_{k=1}^n$ are independent standard normal random variables which are independent of $X_1^n$). Since
\begin{equation}
\lim_{n\rightarrow \infty} \Pr\{Y_{3}^n = 2X_{1}^n + Z_{3}^n \}=1 \label{eqn7Example2}
\end{equation}
by \eqref{probability3Example2}, \eqref{eqn5Example2} and \eqref{eqn4Example2}, it follows that node~1 can use the aforementioned Gaussian codebooks to achieve rate $\frac{1}{2}\log(1+2P)$ for the three-node Gaussian causal relay network, which implies from \eqref{eqn6Example2} that the positive-delay region is strictly smaller than the capacity region for the Gaussian causal relay network for $P>5/4$.
\end{Example}
\medskip
\begin{Remark} \label{remarkIntuition2}
For the three-node Gaussian causal relay network in Example~\ref{exampleCausalRelayNetwork}, the noises in both channels $q_{Y_2|X_1}^{(1)}$ and $q_{Y_3|X_1, X_2, Y_2}^{(2)}$ are correlated as shown in \eqref{probability1Example2} and \eqref{probability2Example2} (or Figure~\ref{GaussianCausalRelay}). If node~2 incurs no delay, node~2 can employ some sort of ``dirty-paper coding" as described above so that noise $Z_{2}^n$ in the signal received by node~3 can be neutralized with high probability (cf.\ \eqref{eqn7Example2}). In contrast, if node~2 incurs a delay, then it cannot use the signal received in a time slot to neutralize the noise incurred on channel $q_{Y_3|X_1, X_2, Y_2}^{(2)}$  in the same time slot, resulting in a lower capacity compared with the case when node~2 incurs no delay.
\end{Remark}

\section{Conclusion} \label{conclusion}
We define the generalized DMN which contains the classical DMN as a special case. In the generalized DMN, some nodes may incur no delay as long as no deadlock occurs. Since every node in the classical DMN incurs a delay, the generalized DMN cannot be modeled by the classical DMN. We prove the cut-set outer bound on the capacity region of the generalized DMN, which subsumes the cut-set bound for the classical DMN.
Then, we investigate the BSC with correlated feedback, which can be modeled as a two-node generalized DMN where one node incurs no delay, and determine the capacity region by proving the tightness of our cut-set bound for this special case.

Next, we establish the cut-set outer bound on the positive-delay region of the generalized DMN. For the BSC with correlated feedback, we show by applying our cut-set bound on the positive-delay region that the positive-delay region is strictly smaller than the capacity region.

Finally, we demonstrate that the causal relay network, which is a generalization of the relay-without-delay channel, is a special case of the generalized DMN. Then, we use our cut-set bound on the capacity region to recover an existing cut-set bound for the causal relay network. In addition, we use our cut-set bound on the positive-delay region to demonstrate that for some Gaussian causal relay network, the positive-delay region is strictly smaller than the capacity region.
\appendix
A proof of Theorem~\ref{thmEquivalentNetwork} is given in this section. The proof involves the following two propositions.
\medskip
\begin{Proposition}\label{propositionCutsetMCLemma*}
Let $(\mathcal{X}_\mathcal{I}, \mathcal{Y}_\mathcal{I}, \alpha, \boldsymbol{\mathcal{S}}, \boldsymbol{\mathcal{G}}, \boldsymbol{q})$ be a DMN. Fix any $(\boldsymbol{1}, n, M_{\mathcal{I}\times\mathcal{I}})$-code on the network and let \linebreak $p_{X_\mathcal{I}, Y_\mathcal{I}}$ denote the probability distribution induced by the code. Then, for each $k\in\{1, 2, \ldots, n\}$ and each $h\in\{1, 2, \ldots, \alpha\}$,
\begin{equation}
\left((W_{\mathcal{I}\times \mathcal{I}}, X_{\mathcal{I}}^{k-1}, Y_{\mathcal{I}}^{k-1},X_{\mathcal{S}_h,k}) \rightarrow X_{\mathcal{S}^{h-1},k} \rightarrow Y_{\mathcal{G}^{h-1},k}\right)_p
\label{cutsetstatement1***}
\end{equation}
forms a Markov Chain.
\end{Proposition}
\begin{IEEEproof}
Let $U^{k-1}=(W_{\mathcal{I}\times \mathcal{I}}, X_{\mathcal{I}}^{k-1}, Y_{\mathcal{I}}^{k-1})$ be the collection of random variables that are generated before the $k^{\text{th}}$ time slot for the $(\boldsymbol{1}, n, M_{\mathcal{I}\times\mathcal{I}})$-code.
Consider the following chain of inequalities for each $k\in\{1, 2, \ldots, n\}$ and each $h\in\{1, 2, \ldots, \alpha\}$:
\begin{align*}
& \!\!\!\! I_{p_{X_\mathcal{I}, Y_\mathcal{I}}}(U^{k-1}, X_{\mathcal{S}_h,k} ; Y_{\mathcal{G}^{h-1},k} | X_{\mathcal{S}^{h-1},k})  \\
&  = \sum_{m=1}^{h-1} I_{p_{X_\mathcal{I}, Y_\mathcal{I}}}(U^{k-1}, X_{\mathcal{S}_h,k} ; Y_{\mathcal{G}_m, k} | X_{\mathcal{S}^{h-1},k}, Y_{\mathcal{G}^{m-1}, k})
\\
& = \sum_{m=1}^{h-1} H_{p_{X_\mathcal{I}, Y_\mathcal{I}}}( Y_{\mathcal{G}_m, k} | X_{\mathcal{S}^{h-1},k}, Y_{\mathcal{G}^{m-1}, k}) -  H_{p_{X_\mathcal{I}, Y_\mathcal{I}}}( Y_{\mathcal{G}_m, k} |U^{k-1},  X_{\mathcal{S}^{h},k}, Y_{\mathcal{G}^{m-1}, k}) \\
& \le  \sum_{m=1}^{h-1} ( H_{p_{X_\mathcal{I}, Y_\mathcal{I}}}( Y_{\mathcal{G}_m, k} | X_{\mathcal{S}^{m},k}, Y_{\mathcal{G}^{m-1}, k}) -  H_{p_{X_\mathcal{I}, Y_\mathcal{I}}}( Y_{\mathcal{G}_m, k} |U^{k-1},  X_{\mathcal{S}^{h},k}, Y_{\mathcal{G}^{m-1}, k}) )
\\
& \stackrel{\text{(a)}}{=}   \sum_{m=1}^{h-1} ( H_{p_{X_\mathcal{I}, Y_\mathcal{I}}}( Y_{\mathcal{G}_m, k} | X_{\mathcal{S}^{m},k}, Y_{\mathcal{G}^{m-1}, k}) -  H_{p_{X_\mathcal{I}, Y_\mathcal{I}}}( Y_{\mathcal{G}_m, k} |U^{k-1},  X_{\mathcal{S}^{m},k}, Y_{\mathcal{G}^{m-1}, k}))
\\
& \stackrel{\text{(b)}}{=} \sum_{m=1}^{h-1} ( H_{p_{X_\mathcal{I}, Y_\mathcal{I}}}( Y_{\mathcal{G}_m,  k} | X_{\mathcal{S}^{m},k}, Y_{\mathcal{G}^{m-1}, k}) - H_{p_{X_\mathcal{I}, Y_\mathcal{I}}}( Y_{\mathcal{G}_m, k} | X_{\mathcal{S}^{m},k}, Y_{\mathcal{G}^{m-1}, k})) \\
&  = 0,
\end{align*}
where
\begin{enumerate}
\item[(a)] follows from Definition~\ref{defCode} that for the $(\boldsymbol{1}, n, M_{\mathcal{I}\times\mathcal{I}})$-code, $X_{\mathcal{I},k}$ is a function of $U^{k-1}$.
\item[(b)] follows from Definition~\ref{defMemoryless} that
\[
\left(U^{k-1} \rightarrow (X_{\mathcal{S}^{m},k}, Y_{\mathcal{G}^{m-1}, k}) \rightarrow Y_{\mathcal{G}_m,  k}\right)_p
\]
 forms a Markov Chain.
    \end{enumerate}
Consequently, $I_{p_{X_\mathcal{I}, Y_\mathcal{I}}}(U^{k-1}, X_{\mathcal{S}_h,k} ; Y_{\mathcal{G}^{h-1},k} | X_{\mathcal{S}^{h-1},k})=0$, which implies that \eqref{cutsetstatement1***} is a Markov Chain.
\end{IEEEproof}
\medskip
\begin{Proposition} \label{PropositionequivalentNetwork}
Let $(\mathcal{X}_\mathcal{I}, \mathcal{Y}_\mathcal{I}, \alpha, \boldsymbol{\mathcal{S}}, \boldsymbol{\mathcal{G}}, \boldsymbol{q})$ be a DMN. For any $(\boldsymbol{1}, n, M_{\mathcal{I}\times\mathcal{I}})$-code on the network, if some $u$, $x_{\mathcal{I}}$ and $y_{\mathcal{I}}$ satisfy
\begin{equation}
\Pr\{U^{k-1} = u, X_{\mathcal{I}, k} = x_\mathcal{I}\}>0 \label{assumption0}
 \end{equation}
 and
 \begin{equation}
 \Pr\{U^{k-1} = u, X_{\mathcal{I}, k} = x_{\mathcal{I}},  Y_{\mathcal{I},k} = y_{\mathcal{I}} \}=0, \label{contradiction1}
  \end{equation}
then there exists some $h\in\{1, 2, \ldots, \alpha\}$ such that
\[
p_{Y_{\mathcal{G}_h} | X_{\mathcal{S}^h}, Y_{\mathcal{G}^{h-1}}}^{(h)}(y_{\mathcal{G}_h} | x_{\mathcal{S}^h}, y_{\mathcal{G}^{h-1}})=0
 \]
 (where $x_{\mathcal{S}^h}$ is a subtuple of $x_{\mathcal{I}}$, and $y_{\mathcal{G}_h}$ and $y_{\mathcal{G}^{h-1}}$ are subtuples of $y_{\mathcal{I}}$).
\end{Proposition}
\begin{IEEEproof}
We prove the proposition by assuming the contrary. Assume
\begin{equation}
p_{Yy_{\mathcal{G}_h} | X_{\mathcal{S}^h}, Y_{\mathcal{G}^{h-1}}}^{(h)}(y_{\mathcal{G}_h} | x_{\mathcal{S}^h}, y_{\mathcal{G}^{h-1}})>0 \label{assumption1}
\end{equation} for all $h\in\{1, 2, \ldots, \alpha\}$.
 We now prove by induction on $h$ that
\begin{equation}
\Pr\{U^{k-1}\! = u, X_{\mathcal{S}^{h},k} \! = x_{\mathcal{S}^{h}}, \! Y_{\mathcal{G}^{h},k}\!=y_{\mathcal{G}^{h}} \}>0 \label{assumption2}
\end{equation}
for each $h\in\{1, 2, \ldots, \alpha\}$. For $h=1$, the LHS of \eqref{assumption2} is
\begin{align}
& \Pr\{U^{k-1}\! = u, X_{\mathcal{S}^{1},k} \! = x_{\mathcal{S}^{1}}, \! Y_{\mathcal{G}^{1},k}\!=y_{\mathcal{G}^{1}} \} \notag \\
& \quad \stackrel{\text(a)}{=} p_{U^{k-1}, X_{\mathcal{S}^1,k}}( u, x_{\mathcal{S}^1}) q_{Y_{\mathcal{G}_1}}^{(1)}(y_{\mathcal{G}_1} | x_{\mathcal{S}^1}) \notag \\
& \quad \stackrel{\text{(b)}}{>} 0 \label{inductionFirstStatement}
\end{align}
where
\begin{enumerate}
\item[(a)] follows from Definition~\ref{defMemoryless}.
\item[(b)] follows from \eqref{assumption0} and \eqref{assumption1}.
\end{enumerate}
If \eqref{assumption2} holds for $h=m$, i.e.,
\begin{equation}
\Pr\{U^{k-1}\! = u, X_{\mathcal{S}^{m},k} \! = x_{\mathcal{S}^{m}}, \! Y_{\mathcal{G}^{m},k}\!=y_{\mathcal{G}^{m}} \}>0, \label{assumption2*}
\end{equation}
 then for $h=m+1$ such that $m+1 \le \alpha$,
\begin{align}
& \Pr\{U^{k-1} = u, X_{\mathcal{S}^{m+1},k} =x_{\mathcal{S}^{m+1}}, Y_{\mathcal{G}^{m+1},k}=y_{\mathcal{G}^{m+1}} \}\notag \\
&\stackrel{\text{(a)}}{=} p_{U^{k-1}, X_{\mathcal{S}^{m+1},k},Y_{\mathcal{G}^{m},k}}( u, x_{\mathcal{S}^{m+1}}, y_{\mathcal{G}^{m}} ) q_{Y_{\mathcal{G}_{m+1}} | X_{\mathcal{S}^{m+1}}, Y_{\mathcal{G}^{m}}}^{(m+1)}(y_{\mathcal{G}_{m+1}} | x_{\mathcal{S}^{m+1}}, y_{\mathcal{G}^{m}})\notag \\
&\stackrel{\text{(b)}}{=}p_{U^{k-1}, X_{\mathcal{S}^{m+1},k}}( u, x_{\mathcal{S}^{m+1}}) p_{Y_{\mathcal{G}^{m},k}|X_{\mathcal{S}^{m},k}}(y_{\mathcal{G}^{m}}|x_{\mathcal{S}^{m}}) q_{Y_{\mathcal{G}_{m+1}} | X_{\mathcal{S}^{m+1}}, Y_{\mathcal{G}^{m}}}^{(m+1)}(y_{\mathcal{G}_{m+1}} | x_{\mathcal{S}^{m+1}}, y_{\mathcal{G}^{m}})\notag \\
 &\stackrel{\text{(c)}}{>} 0, \label{inductionStatement}
\end{align}
where
\begin{enumerate}
\item[(a)] follows from Definition~\ref{defMemoryless}.
\item[(b)] follows from \eqref{cutsetstatement1***} in Lemma~\ref{propositionCutsetMCLemma*}.
\item[(c)] follows from \eqref{assumption0}, \eqref{assumption2*} and \eqref{assumption1}.
\end{enumerate}

For $h=1$, it follows from \eqref{inductionFirstStatement} that \eqref{assumption2} holds. For all $1\le m \le \alpha-1$, it follows from \eqref{assumption2*} and \eqref{inductionStatement} that if \eqref{assumption2} is assumed to be true for $h=m$, then \eqref{assumption2} is also true for $h=m+1$. Consequently, it follows by mathematical induction that \eqref{assumption2} holds for $h=1, 2, \ldots, \alpha$. Since \eqref{assumption2} hold for $h=\alpha$, it follows that
\begin{equation*}
\Pr\{U^{k-1} = u, X_{\mathcal{S}^{\alpha},k}  = x_{\mathcal{S}^{\alpha}}, Y_{\mathcal{G}^{\alpha},k}=y_{\mathcal{G}^{\alpha}} \}>0,
\end{equation*}
which contradicts \eqref{contradiction1}.
\end{IEEEproof}
\medskip
We are now ready to prove Theorem~\ref{thmEquivalentNetwork}.
\begin{IEEEproof}[\textbf{Proof of Theorem~\ref{thmEquivalentNetwork}}]
It suffices to show that for any $(\boldsymbol{1}, n, M_{\mathcal{I}\times\mathcal{I}})$-code, \eqref{memorylessStatement} in Definition~\ref{defMemoryless} is equivalent to
\begin{align}
&  \Pr\{U^{k-1} = u , X_{\mathcal{I}, k} = x_\mathcal{I}, Y_{\mathcal{I}, k} = y_\mathcal{I}\} \notag\\
 & \:= \Pr\{U^{k-1}\! = u , X_{\mathcal{I}, k}\! = x_\mathcal{I}\}\prod_{h=1}^\alpha  q_{Y_{\mathcal{G}_h} | X_{\mathcal{S}^h}, Y_{\mathcal{G}^{h-1}}}{(h)}(y_{\mathcal{G}_h} | x_{\mathcal{S}^h}, y_{\mathcal{G}^{h-1}} ) \label{memorylessStatement2}
\end{align}
for each $k\in\{1, 2, \ldots, n\}$.
Fix a $(\boldsymbol{1}, n, M_{\mathcal{I}\times\mathcal{I}})$-code and a $k\in\{1, 2, \ldots, n\}$. Let $U^{k-1}=(W_{\mathcal{I}\times \mathcal{I}}, X_{\mathcal{I}}^{k-1}, Y_{\mathcal{I}}^{k-1})$ be the collection of random variables that are generated before the $k^{\text{th}}$ time slot.

We first show that \eqref{memorylessStatement} implies \eqref{memorylessStatement2}.
Suppose \eqref{memorylessStatement} holds for each $h\in\{1, 2, \ldots, \alpha\}$. Consider the following three mutually exclusive cases: \smallskip \\ \textbf{Case {$\Pr\{U^{k-1} = u, X_{\mathcal{I}, k} = x_\mathcal{I}\}=0$}:} \\
\indent
Both the LHS and the RHS of \eqref{memorylessStatement2} equal zero.  \smallskip \\
 \textbf{Case {$\Pr\{U^{k-1} = u, X_{\mathcal{I}, k} = x_\mathcal{I}\}>0$} and\\ \text{\hspace{0.1 in}} {$\Pr\{U^{k-1} = u, X_{\mathcal{I}, k} = x_{\mathcal{I}},  Y_{\mathcal{I},k} = y_{\mathcal{I}} \}=0$}:}
\\ \indent
 For this case, the LHS of \eqref{memorylessStatement2} equals zero. By Proposition~\ref{PropositionequivalentNetwork}, there exists some $h\in\{1, 2, \ldots, \alpha\}$ such that $q_{Y_{\mathcal{G}_h} | X_{\mathcal{S}^h}, Y_{\mathcal{G}^{h-1}}}^{(h)}(y_{\mathcal{G}_h} | x_{\mathcal{S}^h}, y_{\mathcal{G}^{h-1}})=0$, which implies that the RHS of \eqref{memorylessStatement2} equals zero. \smallskip \\
 \textbf{Case {$\Pr\{U^{k-1} = u, X_{\mathcal{I}, k} = x_{\mathcal{I}},  Y_{\mathcal{I}, k} = y_{\mathcal{I}}\}>0$}:}
 \\ \indent
For this case,
\begin{align*}
&  \Pr\{U^{k-1} = u, X_{\mathcal{I}, k} = x_\mathcal{I}, Y_{\mathcal{I}, k}  = y_\mathcal{I}\} \\
&\quad = p_{U^{k-1}, X_{\mathcal{I},k}}(u, x_{\mathcal{I}})  p_{Y_{\mathcal{I},k}|U^{k-1},X_{\mathcal{I},k}}(y_{\mathcal{I}} |u , x_{\mathcal{I}})\\
&\quad=  p_{U^{k-1}, X_{\mathcal{I},k}}(u, x_{\mathcal{I}})\prod_{h=1}^\alpha p_{Y_{\mathcal{G}_h,k} |U^{k-1} , X_{\mathcal{I},k}, Y_{\mathcal{G}^{h-1},k}}(y_{\mathcal{G}_h} |u , x_{\mathcal{I}}, y_{\mathcal{G}^{h-1}}) \\
&\quad \stackrel{\text{(a)}}{=} p_{U^{k-1}, X_{\mathcal{I},k}}(u, x_{\mathcal{I}})  \prod_{h=1}^\alpha p_{Y_{\mathcal{G}_h,k} |U^{k-1} , X_{\mathcal{S}^h,k}, Y_{\mathcal{G}^{h-1},k}}(y_{\mathcal{G}_h} |u , x_{\mathcal{S}^h}, y_{\mathcal{G}^{h-1}}) \\
&\quad  \stackrel{\eqref{memorylessStatement}}{=} p_{U^{k-1}, X_{\mathcal{I},k}}(u, x_{\mathcal{I}}) \prod_{h=1}^\alpha  q_{Y_{\mathcal{G}_h} | X_{\mathcal{S}^h}, Y_{\mathcal{G}^{h-1}}}^{(h)}(y_{\mathcal{G}_h} | x_{\mathcal{S}^h}, y_{\mathcal{G}^{h-1}} ), \\
\end{align*}\vspace{-0.4 in}\\
where (a) follows from follows from Definition~\ref{defCode} that for the $(\boldsymbol{1}, n, M_{\mathcal{I}\times\mathcal{I}})$-code, $X_{\mathcal{I},k}$ is a function of $U^{k-1}$.
Therefore, the LHS and the RHS of \eqref{memorylessStatement2} are equal.

Combining the three mutually exclusive cases, we obtain that \eqref{memorylessStatement} implies \eqref{memorylessStatement2}. We now show that \eqref{memorylessStatement2} implies \eqref{memorylessStatement}.
Suppose \eqref{memorylessStatement2} holds. Then for each $h\in\{1, 2, \ldots, \alpha\}$ and each $m\in\{1, 2, \ldots, h\}$,
\begin{align}
&  \Pr\{U^{k-1} = u, X_{\mathcal{S}^h,k} =x_{\mathcal{S}^h}, Y_{\mathcal{G}^{m},k}=y_{\mathcal{G}^{m}} \} \notag \\
&\quad = \sum\limits_{\substack{x_{\mathcal{S}_{h+1}},\ldots, x_{\mathcal{S}_{\alpha}}\\ y_{\mathcal{G}_{m+1}}, \ldots, y_{\mathcal{G}_{\alpha}}}}  p_{U^{k-1}, X_{\mathcal{I}, k}, Y_{\mathcal{I},k}}(u, x_{\mathcal{I}}, y_{\mathcal{I}}) \notag  \\
&\quad \stackrel{\eqref{memorylessStatement2}}{=} \sum\limits_{\substack{x_{\mathcal{S}_{h+1}},\ldots, x_{\mathcal{S}_{\alpha}}\\ y_{\mathcal{G}_{m+1}}, \ldots, y_{\mathcal{G}_{\alpha}}}}  p_{U^{k-1}, X_{\mathcal{I}, k}}(u, x_{\mathcal{I}})\prod_{\ell=1}^\alpha  q_{Y_{\mathcal{G}_\ell} | X_{\mathcal{S}^\ell}, Y_{\mathcal{G}^{\ell-1}}}^{(\ell)}(y_{\mathcal{G}_\ell} | x_{\mathcal{S}^\ell}, y_{\mathcal{G}^{\ell-1}} )\notag   \\
&\quad = \sum_{x_{\mathcal{S}_{h+1}},\ldots, x_{\mathcal{S}_{\alpha}}}  p_{U^{k-1}, X_{\mathcal{I}, k}}(u, x_{\mathcal{I}}) \prod_{\ell=1}^{m}  q_{Y_{\mathcal{G}_\ell} | X_{\mathcal{S}^\ell}, Y_{\mathcal{G}^{\ell-1}}}^{(\ell)}(y_{\mathcal{G}_\ell} | x_{\mathcal{S}^\ell}, y_{\mathcal{G}^{\ell-1}}) \notag \\
&\quad \stackrel{\text{(a)}}{=} p_{U^{k-1}, X_{\mathcal{S}^h, k}}(u, x_{\mathcal{S}^h}) \prod_{\ell=1}^{m}  q_{Y_{\mathcal{G}_\ell} | X_{\mathcal{S}^\ell}, Y_{\mathcal{G}^{\ell-1}}}^{(\ell)}(y_{\mathcal{G}_\ell} | x_{\mathcal{S}^\ell}, y_{\mathcal{G}^{\ell-1}}) \label{equivalentNetworkEqn1}
\end{align}
where (a) follows from the fact that $m\le h$. Then, for each $h\in\{1, 2, \ldots, \alpha\}$, the equality in \eqref{memorylessStatement} can be verified by substituting \eqref{equivalentNetworkEqn1} into the LHS and the RHS.
\end{IEEEproof}

\section*{Acknowledgment}
The authors would like to thank Young-Han Kim for an initial discussion, and Gerhard Kramer for his comment on an earlier version of this paper. The work of Raymond Yeung was partially funded by a grant from the University Grants
Committee of the Hong Kong Special Administrative Region (Project No.\
AoE/E-02/08) and a grant from the Shenzhen Key Laboratory of Network Coding Key
Technology and Application, Shenzhen, China (ZSDY20120619151314964).


\ifCLASSOPTIONcaptionsoff
  \newpage
\fi

\end{document}